\documentclass[preprintnumbers,superscriptaddress,nofootinbib,aps,prd,floatfix]{revtex4}

\usepackage{amsmath,slashed}
\usepackage{graphicx}

\newcommand{\be}{\begin{eqnarray*}}
\newcommand{\ee}{\end{eqnarray*}}

\newcommand{\bee}{\begin{eqnarray}}
\newcommand{\eee}{\end{eqnarray}}
\newcommand{\beeq}{\begin{equation}}
\newcommand{\eeeq}{\end{equation}}

\newcommand{\MET}{$\slashed{E}_T$~}

\newcommand{\pt}{$p_{\rm{T}}$}
\def\spa#1.#2{\left\langle#1#2\right\rangle}
\def\spb#1.#2{\left[#1#2\right]}
\def\lor#1.#2{\left(#1#2\right)}
\def\sand#1.#2.#3{%
\left\langle\smash{#1}{\vphantom1}^{-}\right|{#2}%
\left|\smash{#3}{\vphantom1}^{-}\right\rangle}
\def\sandp#1.#2.#3{%
\left\langle\smash{#1}{\vphantom1}^{-}\right|{#2}%
\left|\smash{#3}{\vphantom1}^{+}\right\rangle}
\def\sandpp#1.#2.#3{%
\left\langle\smash{#1}{\vphantom1}^{+}\right|{#2}%
\left|\smash{#3}{\vphantom1}^{+}\right\rangle}
\def\sandpm#1.#2.#3{%
\left\langle\smash{#1}{\vphantom1}^{+}\right|{#2}%
\left|\smash{#3}{\vphantom1}^{-}\right\rangle}
\def\sandmp#1.#2.#3{%
\left\langle\smash{#1}{\vphantom1}^{-}\right|{#2}%
\left|\smash{#3}{\vphantom1}^{+}\right\rangle}
\def\spab#1.#2.#3{\langle#1|#2|#3]}
\def\spba#1.#2.#3{[#1|#2|#3\rangle}
\def\spaa#1.#2.#3{\langle#1|#2|#3\rangle}
\def\spbb#1.#2.#3{[#1|#2|#3]}
\def\spaxa#1.#2.#3.#4{\langle#1|#2|#3|#4\rangle}
\def\spbxb#1.#2.#3.#4{[#1|#2|#3|#4]}

\DeclareGraphicsExtensions{.pdf,.png,.jpg}

\usepackage{color}

\begin{document} 

\title{Closing up on Dark Sectors at Colliders: from 14 to 100 TeV}

\begin{abstract}
\noindent We investigate the reach of the LHC Run 2 and that of a future circular hadron collider with up to 100 TeV centre of mass energy
for the exploration of potential Dark Matter sectors.  These dark sectors are conveniently and broadly described by simplified models.
The simplified models we consider provide microscopic descriptions of interactions between the Standard Model
partons and the dark sector particles mediated by the four basic types of messenger fields: scalar, 
pseudo-scalar, vector or axial-vector. 
Our analysis extends and updates the previously
available results for the LHC at 8 and 14 TeV to 100 TeV for models with all four messenger types.
We revisit and improve the analysis at 14 TeV, by studying a variety of analysis techniques,
concluding that the most discriminating variables correspond to the missing 
transverse energy and the azimuthal angle between jets in the final state.  
Going to 100 TeV, the limits on simplified models of Dark Matter are enhanced significantly,  in particular for 
heavier mediators and dark sector particles, for which the available phase space at the LHC is restricted. The possibility of a 100 TeV collider provides an unprecedented coverage of the dark sector basic parameters and a unique opportunity to pin down the particle nature of Dark Matter and its interactions with the Standard Model.

\end{abstract}

\author{Philip Harris}

\affiliation{CERN, CH-1211 Geneva 23, Switzerland }
\author{Valentin V. Khoze}
\affiliation{Institute for Particle Physics Phenomenology, Department
  of Physics,\\Durham University, Durham DH1 3LE, United Kingdom}
\author{Michael Spannowsky}
\affiliation{Institute for Particle Physics Phenomenology, Department
  of Physics,\\Durham University, Durham DH1 3LE, United Kingdom}
\author{Ciaran Williams}
\affiliation{Department of Physics, University at Buffalo \\
The State University of New York, 
Buffalo, NY 14260-1500, USA\\
\smallskip
{\tt philip.coleman.harris@cern.ch, valya.khoze@durham.ac.uk, michael.spannowsky@durham.ac.uk, ciaranwi@buffalo.edu}}

\pacs{}
\preprint{IPPP/15/20}
\preprint{DCPT/15/36}

\maketitle

\section{Introduction}
\label{sec:intro}

The data collected by the Planck mission \cite{Planck:2015xua} confirms that, based on the standard model of cosmology, dark matter constitutes nearly 85\% of the total matter content in the universe. With a natural assumption that 
all matter in the universe, dark and visible, is fundamental, dark matter should be described by a microscopic particle theory\footnote{for a review see e.g. \cite{Bertone:2004pz}}. Hence the quest to establish the identity of dark matter, and its fundamental interactions, amounts to one of the most important goals in particle physics.

The observational evidence for dark matter (DM) was established from gravitational effects on visible matter.
However, the Standard Model (SM) of particle physics does not contain any viable DM 
candidates. In this way dark matter provides us with arguably the strongest  experimental evidence for the 
existence of physics beyond the Standard Model. The observation of non-gravitational interactions of 
DM with visible matter could be crucial in discovering extensions of known fundamental theories. 
Since the experimental evidence for the existence of DM is currently only gravitational, at present we have no definitive 
microscopic understanding of dark matter. For these reasons the search for DM has escalated in recent years with direct detection, indirect detection, and most recently
high energy collider searches combining to restrict the range of parameter space for dark matter models.

Presently the frontier of high-energy physics corresponds to the newly upgraded LHC, which is currently operating at $\sqrt{s}=13$ TeV, with 
a planned increase to $\sqrt{s}=14$ TeV in the near future. However, it is certainly plausible that the scale of new physics is out of range 
of the LHC's reach. Broadly speaking, analyses of Run 1 data are able to constrain new physics approximately up to $2 $ TeV. The higher 
operating energy at Run 2 will certainly increase these bounds, but not by more than an order of magnitude. Therefore if new physics arises first 
at scales greater than those probed by the LHC, it may not be possible to infer its existence with the current generation of machines. 
For this reason, the community is beginning to consider the potential for the successor to the LHC~\cite{FCC}.
The next machine, will have to significantly extend the reach of the LHC, and as such, requires a large increase in operating energy. Current 
proposals set 100 TeV as the target center of mass energy. Since a hadron machine of this energy mandates a circular design the 
putative machine is usually referred to as the Future Circular Collider (FCC). Such a machine would be a formidable tool in the 
quest to discover new physics.

\medskip

The main motivation of this paper is to investigate the reach of the FCC with a 100 TeV centre of mass energy for dark matter searches and 
ideally potential discoveries. In parallel with our earlier work \cite{Harris:2014hga} which analyzed the 
limits of the LHC at 8 TeV and its DM discovery potential at 14 TeV, we aim for a general model-independent characterisation 
of dark sectors with as few free parameters as possible, but without resorting to an effective field theory description.
To achieve this we employ the simplified model approach which provides a microscopic QFT description of a minimal
set of interactions between the Standard Model partons and the dark sector particles. 
These interactions are mediated by a complete set of four basic types of messenger fields, i.e. the scalar, 
pseudo-scalar, vector and axial-vector.  One naturally expects that the FCC should perform significantly better 
than the LHC for heavy mediators and heavy dark matter particles, the primary aim of this work is to seek to quantify this improvement.
Our work is another step in the emerging program of DM studies at future colliders in the 100 TeV range,
\cite{Low:2014cba,Bramante:2014tba,Xiang:2015lfa}. Related studies using simplified models for constraining dark sectors at the LHC include 
 Refs.~\cite{Buckley:2014fba,Abdallah:2014hon,Malik:2014ggr,Buchmueller:2014yoa,Haisch:2015ioa,Chala:2015ama,
 Khoze:2015sra,Buchmueller:2015eea,Fan:2015sza}, and we also refer the reader to the recent summaries  \cite{Abdallah:2015ter,Abercrombie:2015wmb} and references therein.
 
In DM searches at hadron colliders, the putative dark particles are pair-produced in collisions of the visible sector particles -- 
the Standard Model quarks and gluons. In the set-up we study here, there are no direct interactions between 
the SM sector and the dark matter particles. Instead these interactions are mediated by an intermediate degree of freedom --
the mediator field. In general, one can expect four types of mediators, 
scalar $S$, pseudo-scalar $P$, vector $Z'$ or axial-vector $Z''$. The corresponding four classes of simplified models 
describing elementary interactions of these four mediators with the SM quarks and with the dark sector fermions $\chi$ 
are 
\begin{align}
\label{eq:LS} 
\mathcal{L}_{\mathrm{scalar}}&\supset\, -\,\frac{1}{2}m_{\rm MED}^2 S^2 - g_{\rm DM}  S \, \bar{\chi}\chi
 - \sum_q g_{SM}^q S \, \bar{q}q  - m_{\rm DM} \bar{\chi}\chi \,,
 \\
 \label{eq:LP} 
\mathcal{L}_{\rm{pseudo-scalar}}&\supset\, -\,\frac{1}{2}m_{\rm MED}^2 P^2 - i g_{\rm DM}  P \, \bar{\chi} \gamma^5\chi
 -\sum_q  i g_{SM}^q  P \, \bar{q}  \gamma^5q  - m_{\rm DM} \bar{\chi}\chi\,,
 \\
 \label{eq:LV} 
\mathcal{L}_{\mathrm{vector}}&\supset \, \frac{1}{2}m_{\rm MED}^2 Z'_{\mu} Z'^{\mu} - g_{\rm DM}Z'_{\mu} \bar{\chi}\gamma^{\mu}\chi -\sum_q g_{SM}^q Z'_{\mu} \bar{q}\gamma^{\mu}q - m_{\rm DM} \bar{\chi}\chi\,,
 \\
 \label{eq:LA} 
\mathcal{L}_{\rm{axial}}&\supset\,  \frac{1}{2}m_{\rm MED}^2 Z''_{\mu} Z''^{\mu} - g_{\rm DM} Z''_{\mu} \bar{\chi}\gamma^{\mu}\gamma^5\chi -\sum_q g_{SM}^q Z''_{\mu} \bar{q}\gamma^{\mu}\gamma^5q - m_{\rm DM} \bar{\chi}\chi\,.
\end{align}
The coupling constant $g_{\rm DM}$ characterizes the interactions of the messengers with the dark sector particles,
which for simplicity we take to be Dirac fermions $\chi$, $\bar{\chi}$, the case of scalar DM particles is a straightforward extension 
of these results. 

The coupling constants linking the messengers to the SM quarks are collectively described by $g^q_{\rm SM}$,
\begin{eqnarray}
{\rm scalar \,\,\& \,\, pseudo-scalar\, messengers:} && \quad g_{\rm SM}^q \equiv\,  g_q\, y_q\,=\, g_q\, \frac{m_q}{v}\,, 
\label{eq:gdef}\\
{\rm vector \,\,\,\& \,\, axial-vector\, \, messengers:} && \quad g_{\rm SM}^q \, =\, g_{\rm SM}\,.
\label{eq:gdef2}
\end{eqnarray}
For scalar and pseudo-scalar messengers the couplings to quarks are taken to be proportional to the corresponding Higgs Yukawa couplings, $y_q$ as in models with minimal flavour violation \cite{D'Ambrosio:2002ex}, and we keep the scaling 
$g_q$ flavour-universal for all quarks.
For axial and vector mediators  $g_{\rm SM}$ is a gauge coupling in the dark sector which we also
take to be flavour universal.
The coupling parameters which we can vary are thus $g_{\rm DM}$ plus either $g_q$ or $g_{\rm SM}$, the latter choice depending on the messengers.\footnote{In Ref.~\cite{Harris:2014hga} we have parametrised $g_{\rm DM}$ for (pseudo-)scalar
messengers as $g_{\rm DM}\,=\, g_\chi\, m_{\rm DM}/v$ to look symmetric w.r.t. \eqref{eq:gdef}, and have treated 
$g_\chi$ as a free parameter. Here we do not impose this requirement and leave $g_{\rm DM}$ as the free parameter.}

In general, the simplified model description of the dark sector is characterised by five parameters:
the mediator mass $m_{\rm MED}$, the mediator width 
$\Gamma_{\rm MED}$, the dark particle mass $m_{\rm DM}$, and the mediator-SM and the mediator-Dark sector couplings, 
$g_{\rm SM}$, $g_{\rm DM}$. Out of these, the mediator width $\Gamma_{\rm MED}$,
does not appear explicitly in the simplified model Lagrangians \eqref{eq:LS}-\eqref{eq:LA} and should be specified separately.
$\Gamma_{\rm MED}$ accounts for the allowed decay modes of a given mediator particle into other particles from the visible 
and the dark sector.
In a complete theory, $\Gamma_{\rm MED}$ can be computed from its Lagrangian, but in a simplified model we can instead determine only the
so-called minimal width $\Gamma_{\rm MED, min}$, i.e. the mediator width computed using the mediator interactions with the SM quarks 
and the $\bar{\chi}$, $\chi$ DM particles defined in Eqs.~\eqref{eq:LS}-\eqref{eq:LA}. 
Importantly  $\Gamma_{\rm MED, min}$ does not take into account the possibility of the mediator to decay into
e.g. other particles of the dark sector, beyond $\bar{\chi}$, $\chi$, which would increase the value of $\Gamma_{\rm DM}$.
In Ref.~\cite{Harris:2014hga} we have investigated the role of $\Gamma_{\rm MED}$ as an independent parameter in the simplified models
characterisation of dark sectors by using a simple grid for $\Gamma_{\rm DM}=\{1,2,5,10\}\times \Gamma_{\rm MED, min}$.

Here we will not repeat this analysis, referring  to \cite{Harris:2014hga} for a more general discussion of $\Gamma_{\rm MED}$;
we will instead adopt a reduced simplified description where the width is set to its minimal computed
 value $\Gamma_{\rm MED, min}$ which amounts to larger signal cross-sections 
 (we will also check that $ \Gamma_{\rm MED, min} < m_{\rm MED}/2$). For our simplified models we have
\begin{equation}
\label{eq:GVA}
\Gamma_{\rm MED, min}\,=\, \Gamma_{\chi\overline{\chi}} \,+ \,\sum_{i=1}^{N_f} N_c\, \Gamma_{q_i\overline{q}_i}  
\end{equation}
where $\Gamma_{\chi\overline{\chi}}$ is the mediator decay rate into two DM fermions, and the sum  is over the SM quark flavours.
Depending on the mediator mass, decays to top quarks may or may not be open i.e. $m_{\rm MED}$ should
 be $> 2m_t$ for an open decay. The partial decay widths of vector, Axial-vector, scalar and pseudo-scalar mediators into fermions are
 given by,
\begin{eqnarray}
\label{eq:GV}
\Gamma^{V}_{f\overline{f}} &=& \frac{g^2_{f}(m_{\rm MED}^2+2m_f^2)}{12 \pi m_{\rm MED}}\sqrt{1-\frac{4m_f^2}{m_{\rm MED}^2}} \quad ,
\quad
\Gamma^{A}_{f\overline{f}} \,=\, \frac{g^2_{f}(m_{\rm MED}^2-4m_f^2)}{12 \pi m_{\rm MED}}\sqrt{1-\frac{4m_f^2}{m_{\rm MED}^2}} \\
\label{eq:GS}
\Gamma^{S}_{f\overline{f}} &=& \frac{g^2_{f}}{8\pi }\,m_{\rm MED}\,\left(1-\frac{4m_f^2}{m_{\rm MED}^2}\right)^\frac{3}{2}
\quad ,\quad
\Gamma^{P}_{f\overline{f}} \,=\, \frac{g^2_{f}}{8\pi}\,m_{\rm MED}\,\left(1-\frac{4m_f^2}{m_{\rm MED}^2}\right)^\frac{1}{2}
\end{eqnarray}
where $m_f$ denotes masses of either SM quarks $q$ or DM fermions $\chi$ and the coupling constant $g_{f}$ denotes
either $g_{\rm SM}$ or $g_{\rm DM}$.

\begin{center}
\begin{figure}[h,t]
\includegraphics[width=0.3\textwidth]{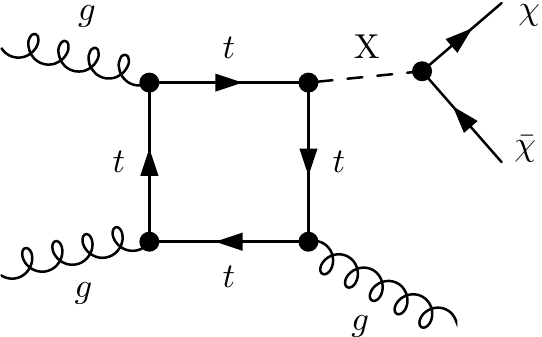} \hspace{0.5cm} 
\includegraphics[width=0.28\textwidth]{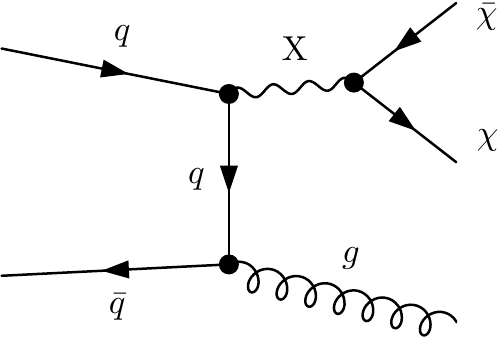}
\caption{Representative Feynman diagrams for gluon and quark induced mono-jet plus MET processes. The mediator X can be a scalar, pseudo-scalar, vector or axial-vector particle. The gluon fusion process involves the heavy quark loop which we compute in the microscopic theory, while the quark-anti-quark annihilation is a tree-level process at leading order.
}
\label{fig:feyn}
\end{figure}
\end{center}

In this paper we will 
focus on jets plus missing energy searches, generalizing our earlier results \cite{Harris:2014hga} 
from the 8-14 TeV to 100 TeV colliders. The analysis in \cite{Harris:2014hga} 
was based on a mono-jet plus missing transverse energy (MET, or \MET) signature -- a popular choice used in searches for new physics including supersymmetry, extra dimensions and dark matter at the Tevatron and the LHC
\cite{Abazov:2003gp,Aaltonen:2012jb, Chatrchyan:2012me, ATLASMONO, Khachatryan:2014rra,Diehl:2014dda,Feng:2005gj,Cao:2009uw,Beltran:2010ww,Goodman:2010yf,Goodman:2010ku,Fox:2011pm,Haisch:2012kf}.
Here we will update the experimental analysis techniques to take into account both the leading and trailing jets in the final state to present a more realistic idea of potential limits. Hence we will update our LHC 14 results  \cite{Harris:2014hga}  accordingly to provide a fair benchmark.

Depending on the choice for the mediator field different production mechanisms will contribute. 
For vectors and axial-vectors the dominant mechanism is the
quark-antiquark annihilation at tree-level.  For scalars and pseudo-scalars on the other hand, 
the loop-level gluon fusion processes are more relevant. The representative Feynman diagrams for both channels are
shown in Fig.~\ref{fig:feyn}. In comparing DM collider searches with direct and indirect detection experiments it is important to keep in mind that 
our collider processes and limits continue to be applicable for discovery of any dark sector particles escaping the detector.
Hence dark particles produced at colliders do not have to be the cosmologically stable dark matter.

\medskip

Finally we would like to comment on the possible origin and the UV consistency of the simplified models \eqref{eq:LS}-\eqref{eq:LA}.
The scalar and pseudo-scalar messenger fields in our simplified models \eqref{eq:LS}-\eqref{eq:LP}
are singlets under the Standard Model. How can this be reconciled with the fact that they are supposed to be Higgs-like, with the Higgs being an $SU(2)_L$ doublet? In fact, the simplified models \eqref{eq:LS}-\eqref{eq:LP} can arise from two types of the 
more fundamental theories. The simplest theories of the first type are the two-Higgs-doublet models \cite{Branco:2011iw}.
In this case the mediators would originate from the second Higgs doublet. The other type of models
giving rise to our simplified models are even simpler in the sense that scalar mediators (and the dark sector particles 
they are coupled to) can be genuinely neutral under the SM but mix with the neutral component of the Higgs.
Following the Higgs discovery there is a renewed interest in the literature in Higgs portal models 
\cite{Silveira:1985rk, Schabinger:2005ei, Patt:2006fw, Englert:2011yb}
where the scalar mediators apartonsre SM-singlets but the SM Higgs $h$ interacts with them via the interaction,
$\lambda_{\rm hp} |H|^2 |\Phi|^2$. When $\Phi$ and $H$ both develop a VEV, mass mixing occurs, and after
transforming into the mass eigenstate basis, one finds two scalar resonances $h_1$ and $h_2$, both of which interact with the Standard Model and the dark sector, with the $h_1$ state identified with the SM Higgs and $h_2$ being the scalar mediator.
These models provide a direct connection of the dark sector with Higgs physics and can link 
the origin of the electroweak and the DM scales~\cite{Englert:2013gz,Hambye:2013dgv,Carone:2013wla,Khoze:2014xha}.
The simplified dark sector models with vector and axial-vector mediators in Eqs.~\eqref{eq:LV}-\eqref{eq:LA}
can also be derived from appropriate first-principles 
theories.  Since the mediators are spin-one particles, these UV models would necessarily require the mediators 
to be gauge fields and the DM to be charged under these gauge transformations.
A classification of anomaly-free 
extensions of the Standard Model  Abelian $U(1)'$ factor was given in \cite{Carena:2004xs}
and can be used for constructing an example of a consistent gauge-invariant vector and axial theories of the type \eqref{eq:LA}.

\medskip

This paper proceeds as follows, in section~\ref{sec:setup} we discuss the theoretical and experimental setup we used for our analysis. 
The results of our analysis are presented in section~\ref{sec:results}, and our conclusions are presented in section~\ref{sec:conc}.

\section{Theoretical and experimental setup} 
\label{sec:setup} 

\subsection{Collider design and Monte Carlo generation}
\label{sec:detmc}

At this moment, the exact nature of the FCC is unclear. This level of uncertainty includes
fundamental issues, like the operating center of mass energy of the machine, and technical 
details including, for instance, the capabilities of future detectors. This makes robust predictions 
impossible at the present time, instead it is more interesting to study what is potentially feasible at 
future colliders, given a modest (and conservative) set of assumptions. 

We begin by discussing the straightforward benefits associated with higher energy machines in relation to 
DM searches. Clearly with larger center of mass energies, heavier mediators can be produced on-shell, allowing 
for significant enhancements for heavy mediators compared to a 14 TeV analysis.  Secondly there are effects 
due to the parton distribution functions (PDFs). A typical 14 TeV DM analysis naturally focusses on the region of
high missing transverse energy, which corresponds to larger values of the partonic center of mass energy ($\hat{s}$). This region of phase space 
is suppressed by the PDFs, and therefore additionally damps the differential cross section, beyond the simple phase space/matrix element 
falloff. However, at higher energies the same region of phase space now corresponds to much smaller $\hat{s}/s$ values, and as a result
the PDF suppression is no longer as severe. Therefore, even for mass scales which are currently probed at the LHC, we expect 
a higher energy machine to be able to improve upon the bounds. 

A major concern for the simulation of analyses at an FCC, is the modelling of the theoretical event generation. It is clear that 
a huge amount of work will be needed over the coming decades to ensure that the Monte Carlo simulations provide a
reasonable description of signal and background processes~\cite{Avetisyan:2013onh}. We do not attempt this huge feat in this paper, but instead 
list the concerns regarding existing tools, and their potential impact. 

\begin{center}
\begin{figure}[t]
\includegraphics[width=0.45\textwidth]{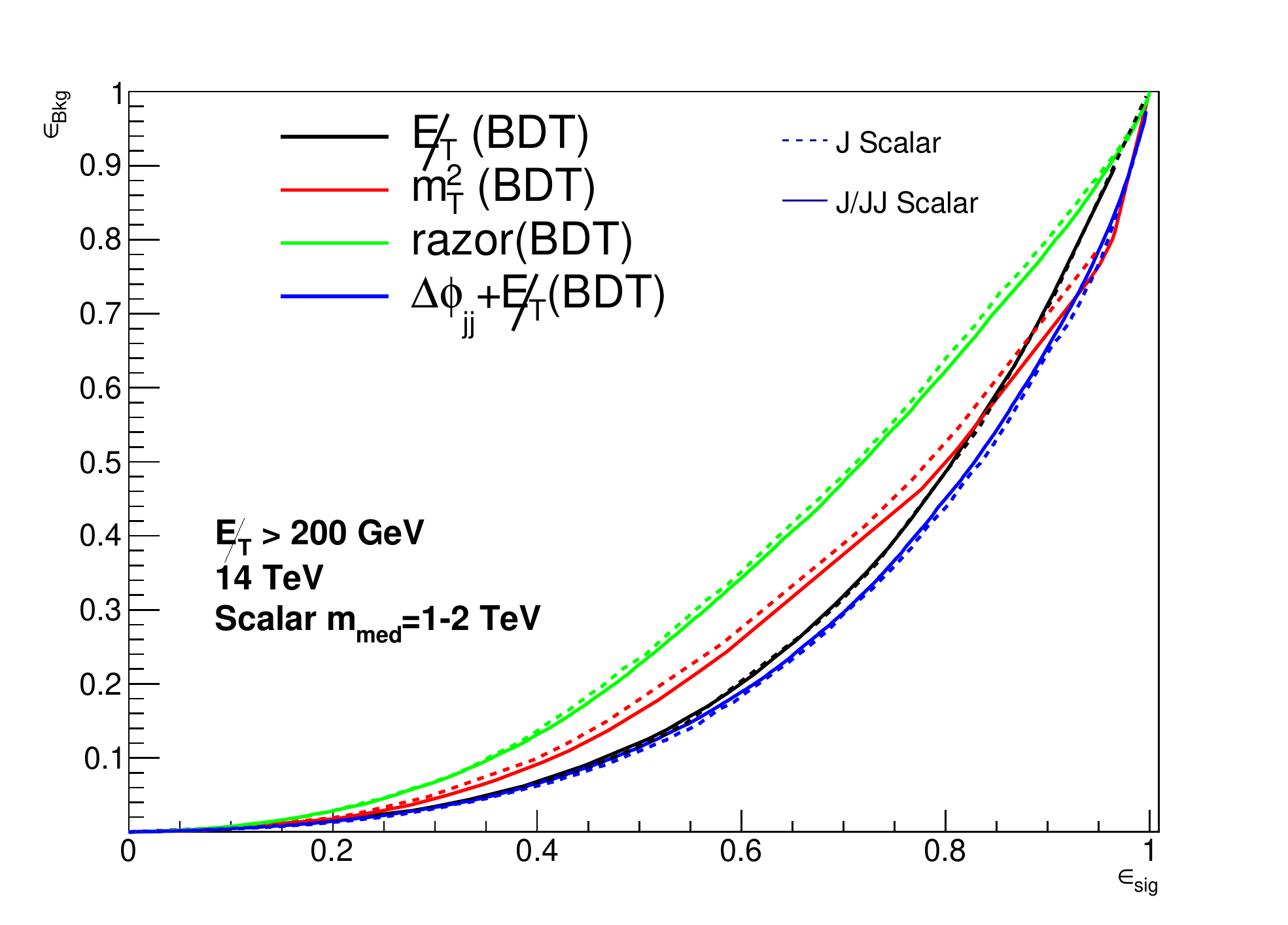}
\includegraphics[width=0.45\textwidth]{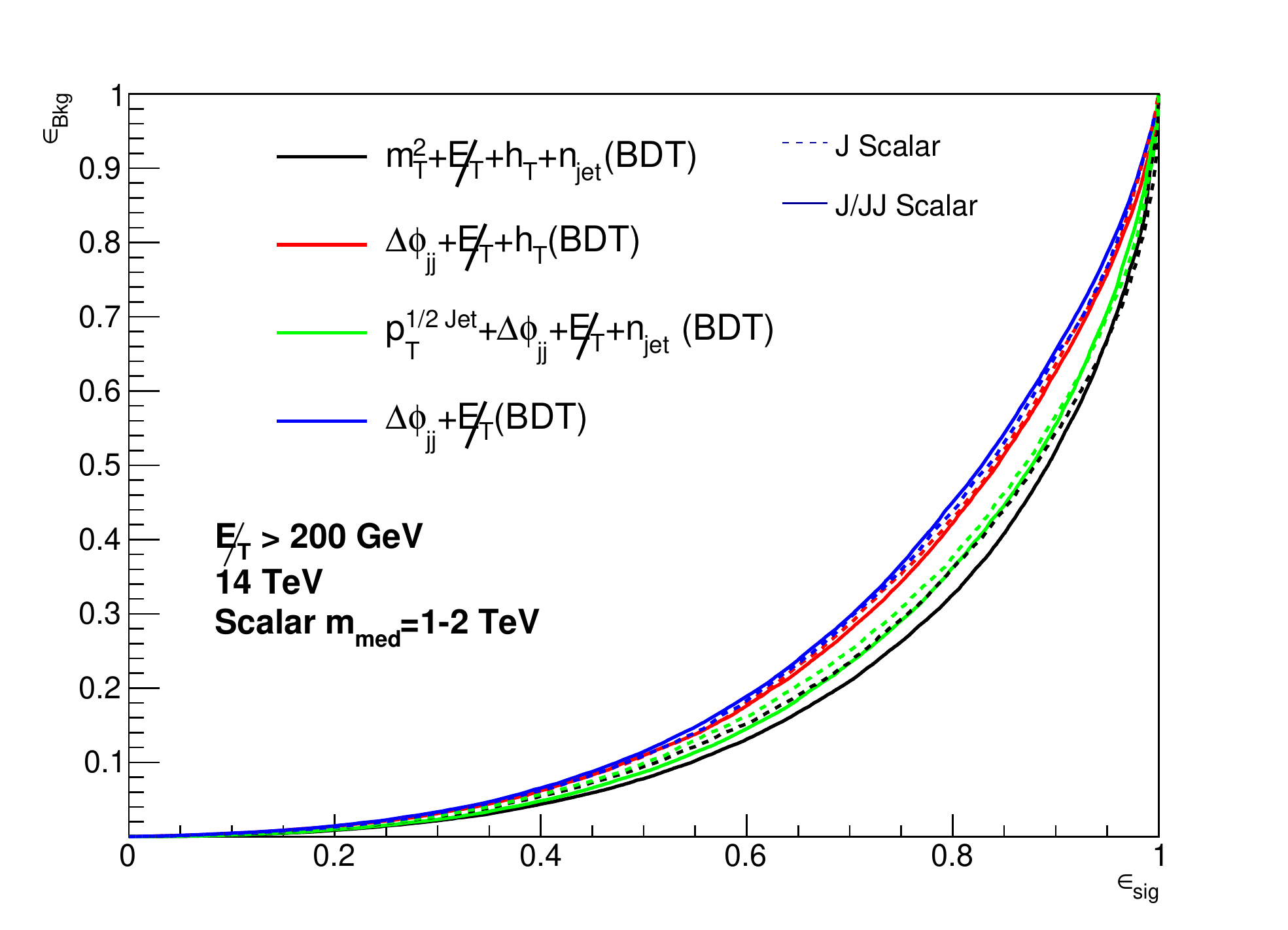}
\caption{Reconstruction efficiency for signal $\epsilon_{\rm sig}$ and background $\epsilon_{\rm bkg}$ as
  functions of individual kinematic variables at $14$ TeV with 1~ab$^{-1}$ luminosity for scalar mediator models using
single  variables as well as the combinations of variables, as indicated. The dashed lines
depict single jet signal MC, while the solid lines correspond to merged samples with up to two jets.}
\label{fig:roc}
\end{figure}
\end{center}

\begin{center}
\begin{figure}[t]
\includegraphics[width=0.5\textwidth]{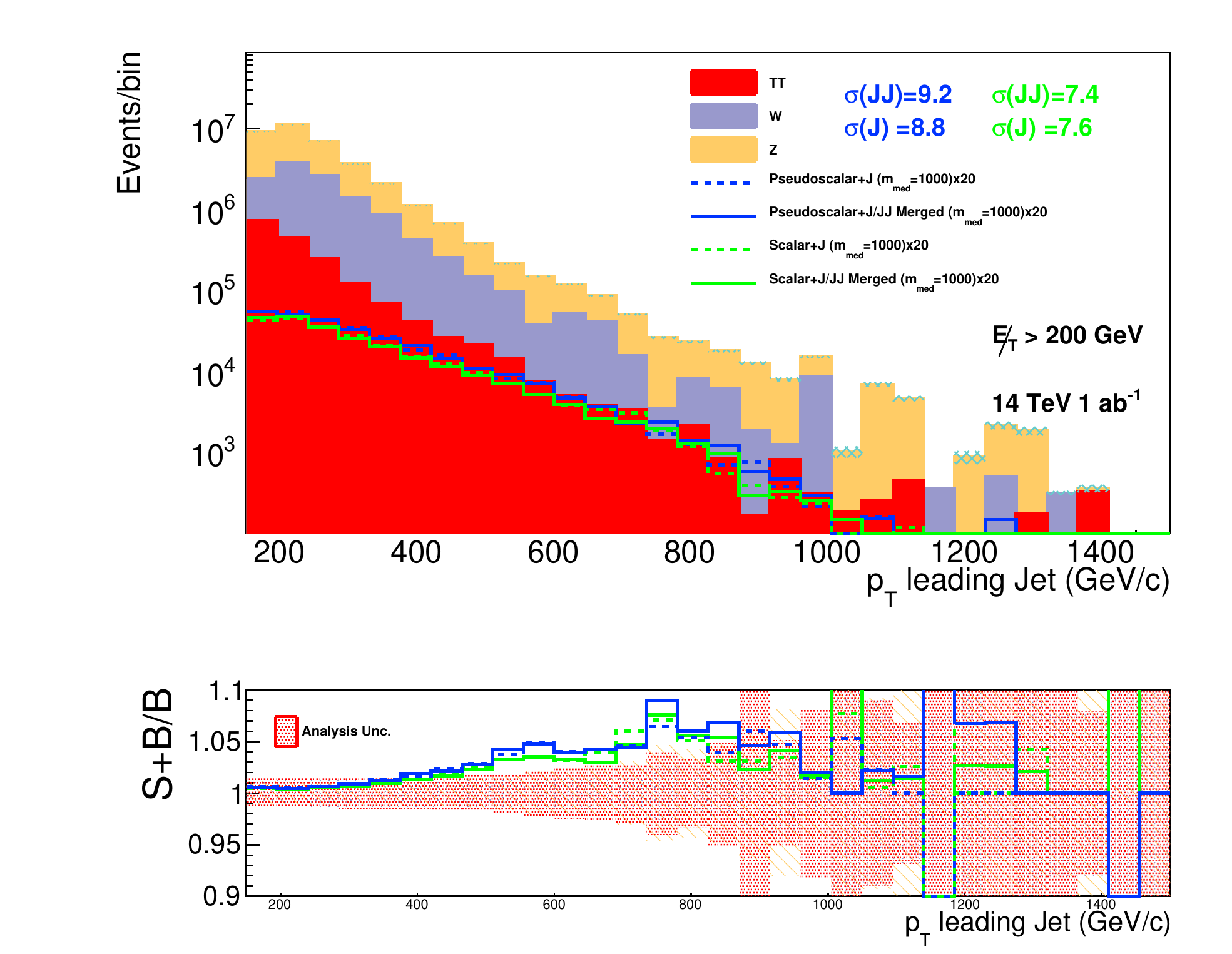} \hskip-0.5cm
\includegraphics[width=0.5\textwidth]{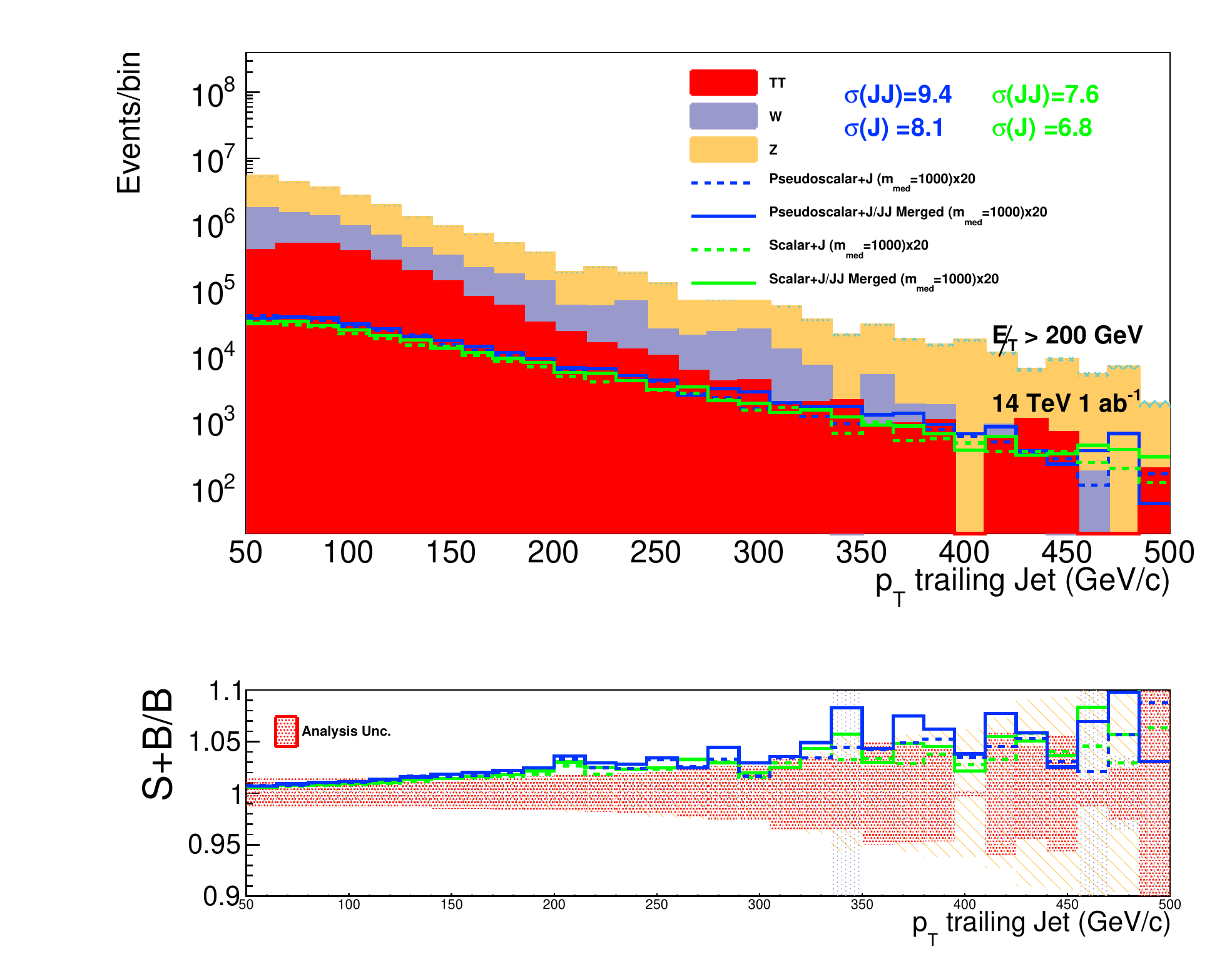}\\
\includegraphics[width=0.5\textwidth]{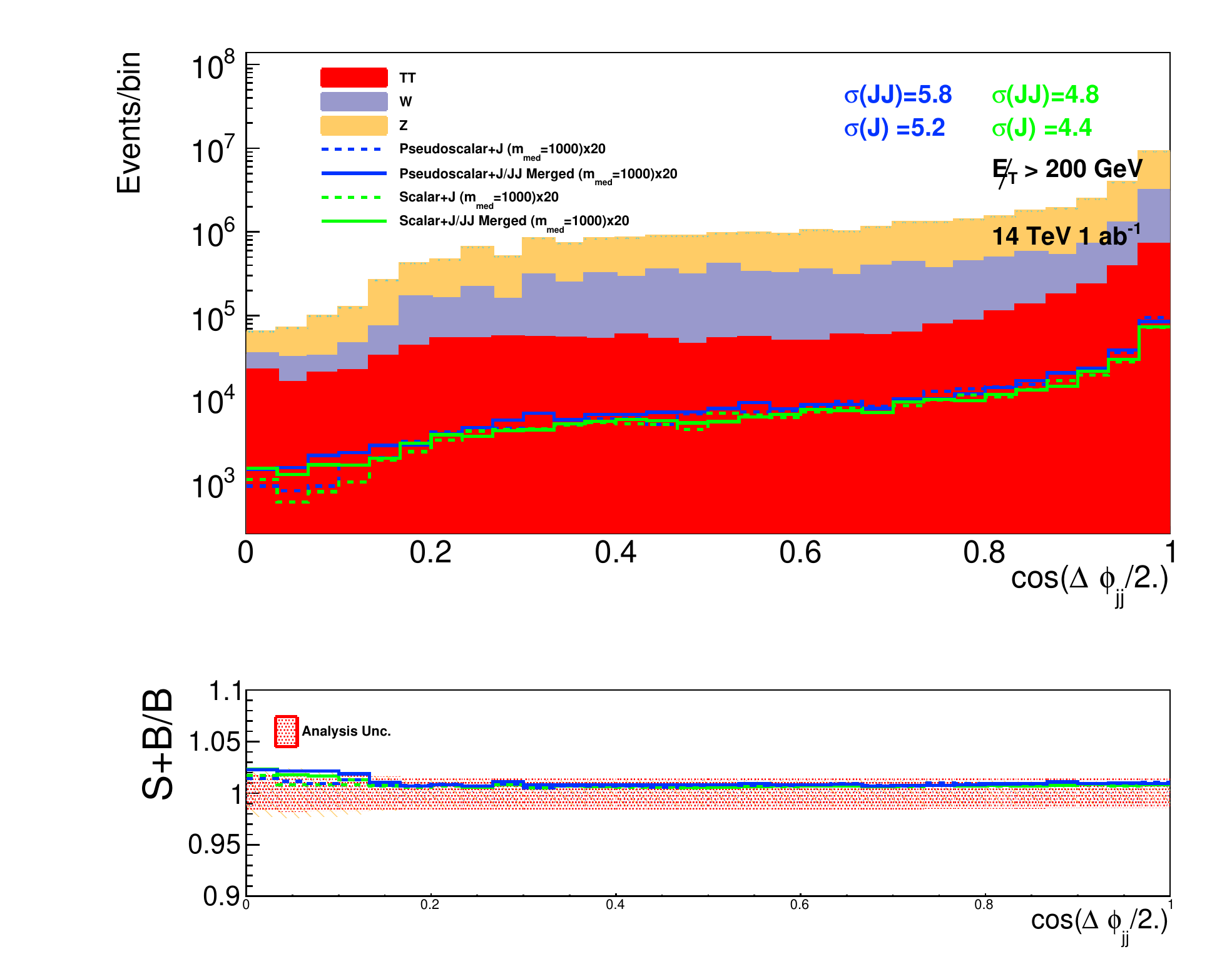} \hskip-0.5cm
\includegraphics[width=0.5\textwidth]{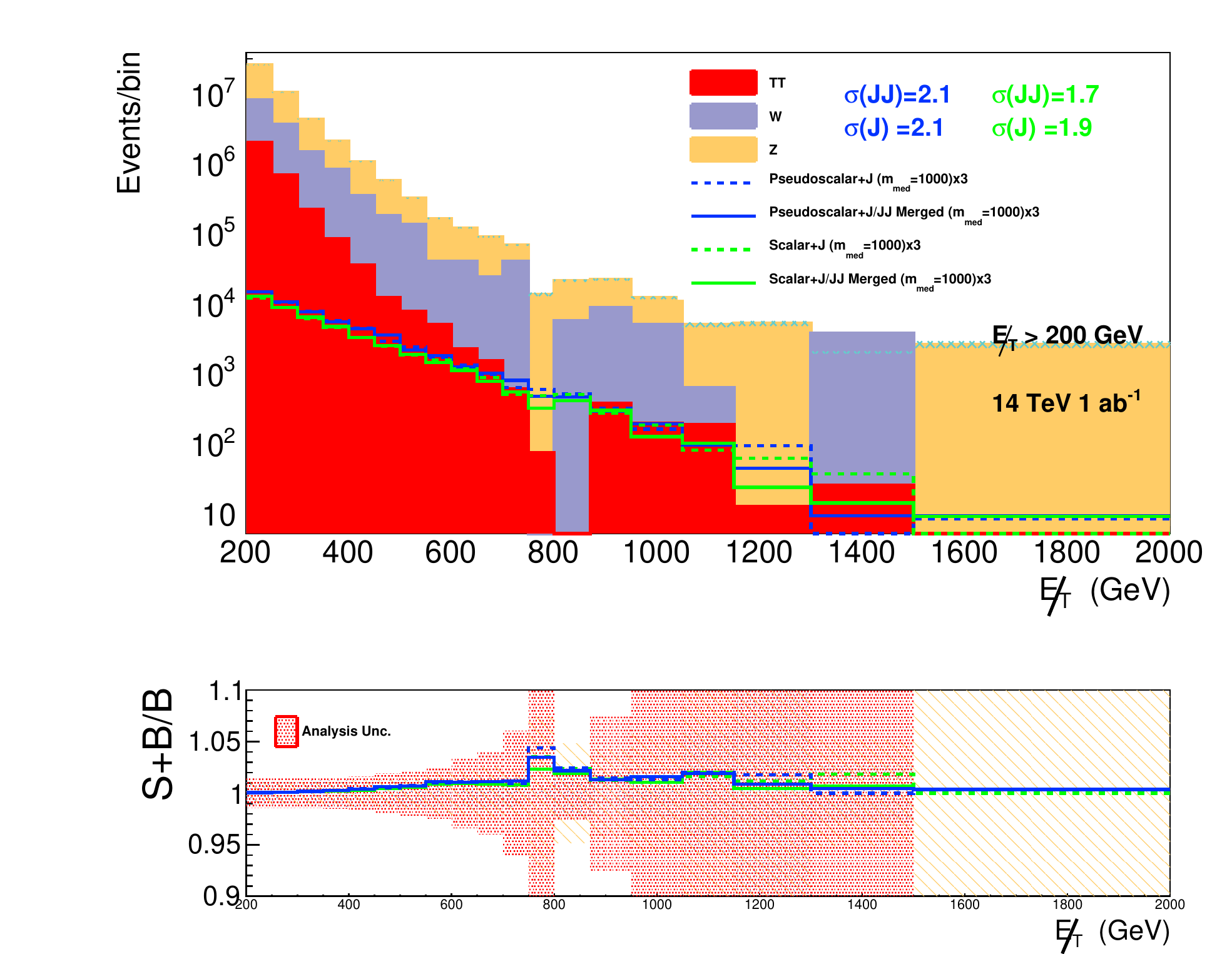}
\caption{Kinematic distributions for signal scalar and mediator models and the SM backgrounds at $14$ TeV 
assuming 1~ab$^{-1}$ of integrated luminosity. We show four kinematic 
variables: $p_T$ of the leading jet, $p_T$ of the trailing jet, the azimuthal angle between the jets $\cos (\Delta \phi_{jj})$, 
and the missing energy \MET. Ratios of $(S+B)/B$ are shown for each observable. The red bands indicate the uncertainties on the background distributions. The accordingly color-coded numbers for sig(JJ) and sig(J) give the statistical significance to disfavour the presence of the signal using the $\mathrm{CL}_s$ method. }
\label{fig:kin14TeVscal}
\end{figure}
\end{center}

\begin{center}
\begin{figure}[t]
\includegraphics[width=0.5\textwidth]{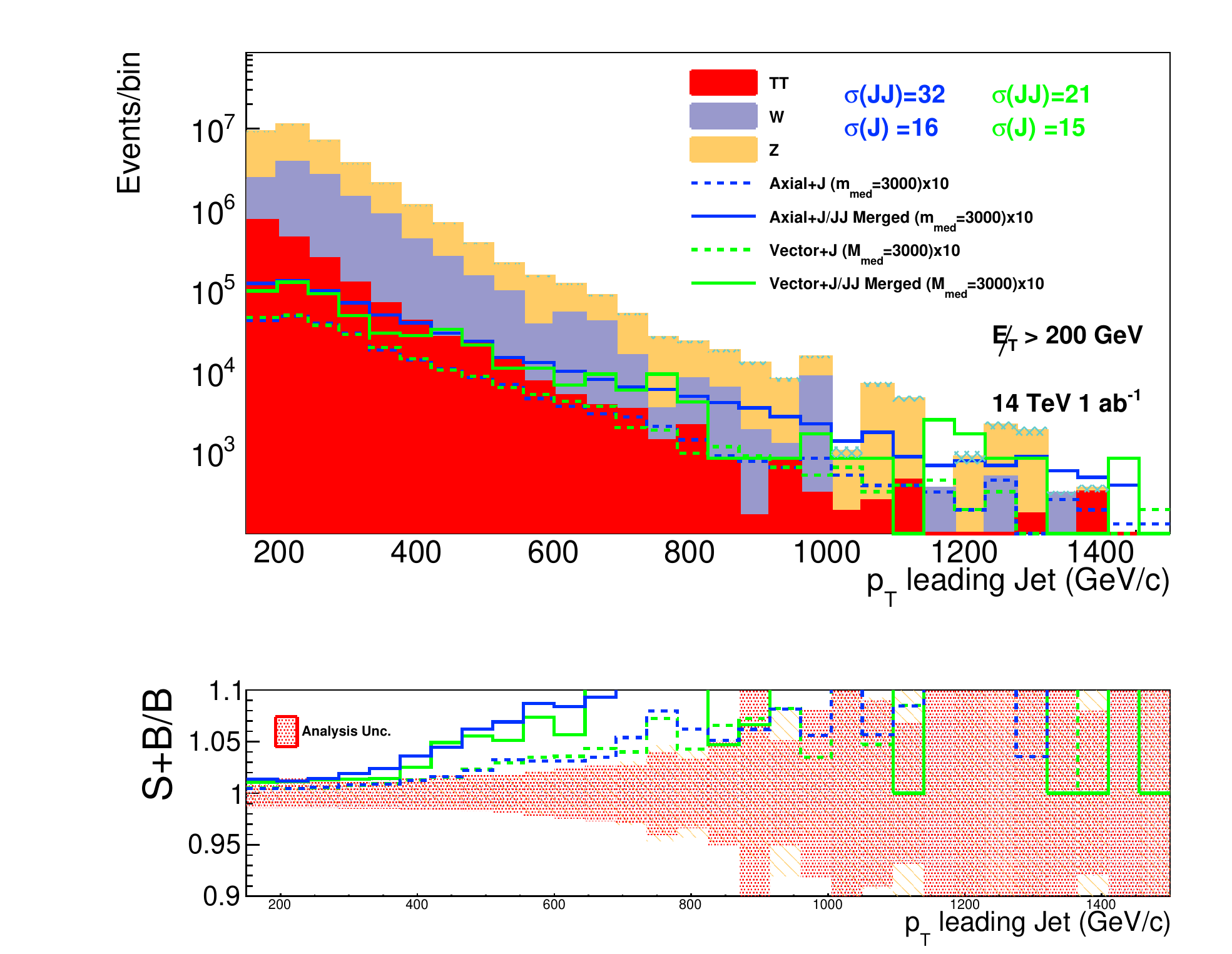} \hskip-0.5cm
\includegraphics[width=0.5\textwidth]{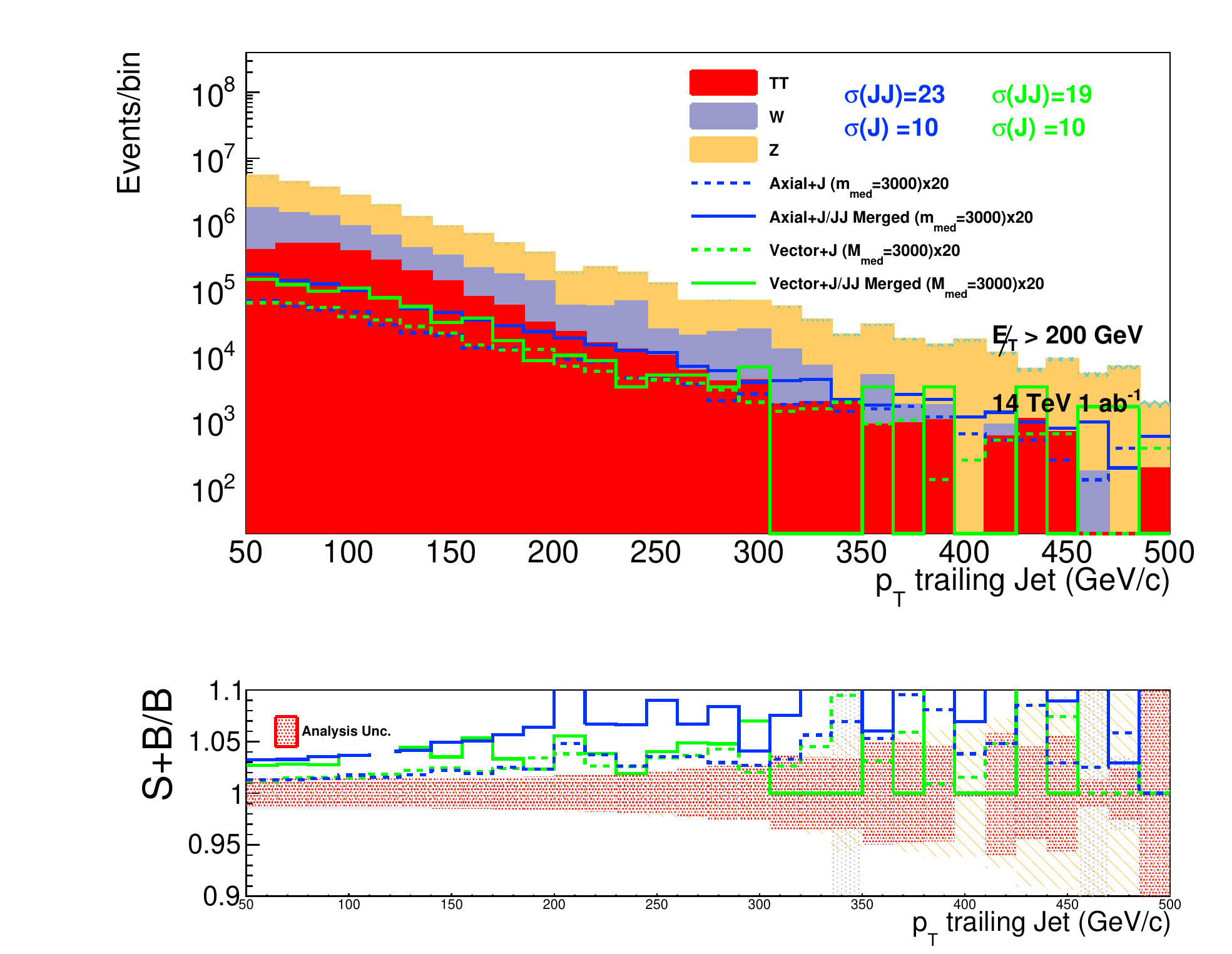}\\
\includegraphics[width=0.5\textwidth]{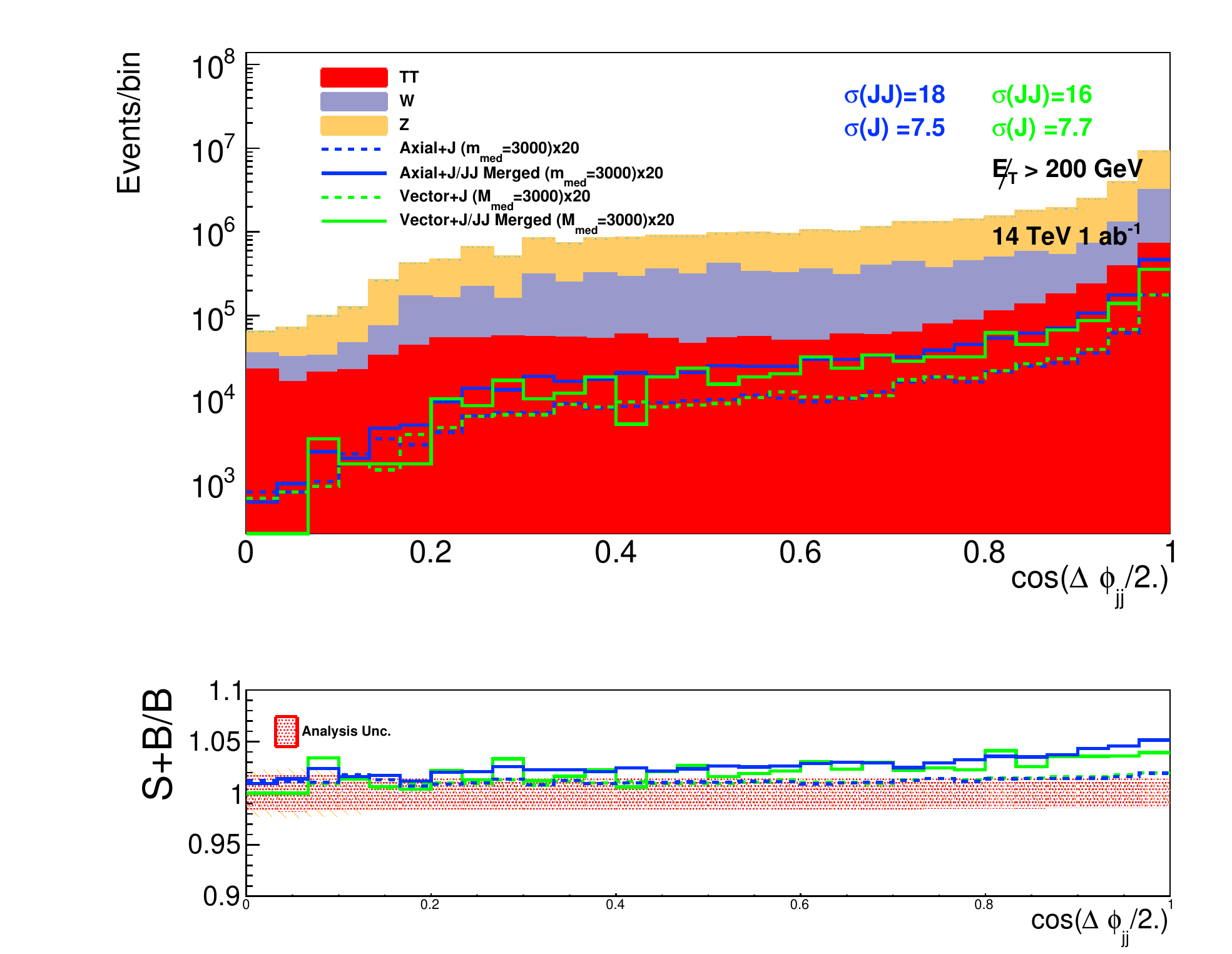} \hskip-0.5cm
\includegraphics[width=0.5\textwidth]{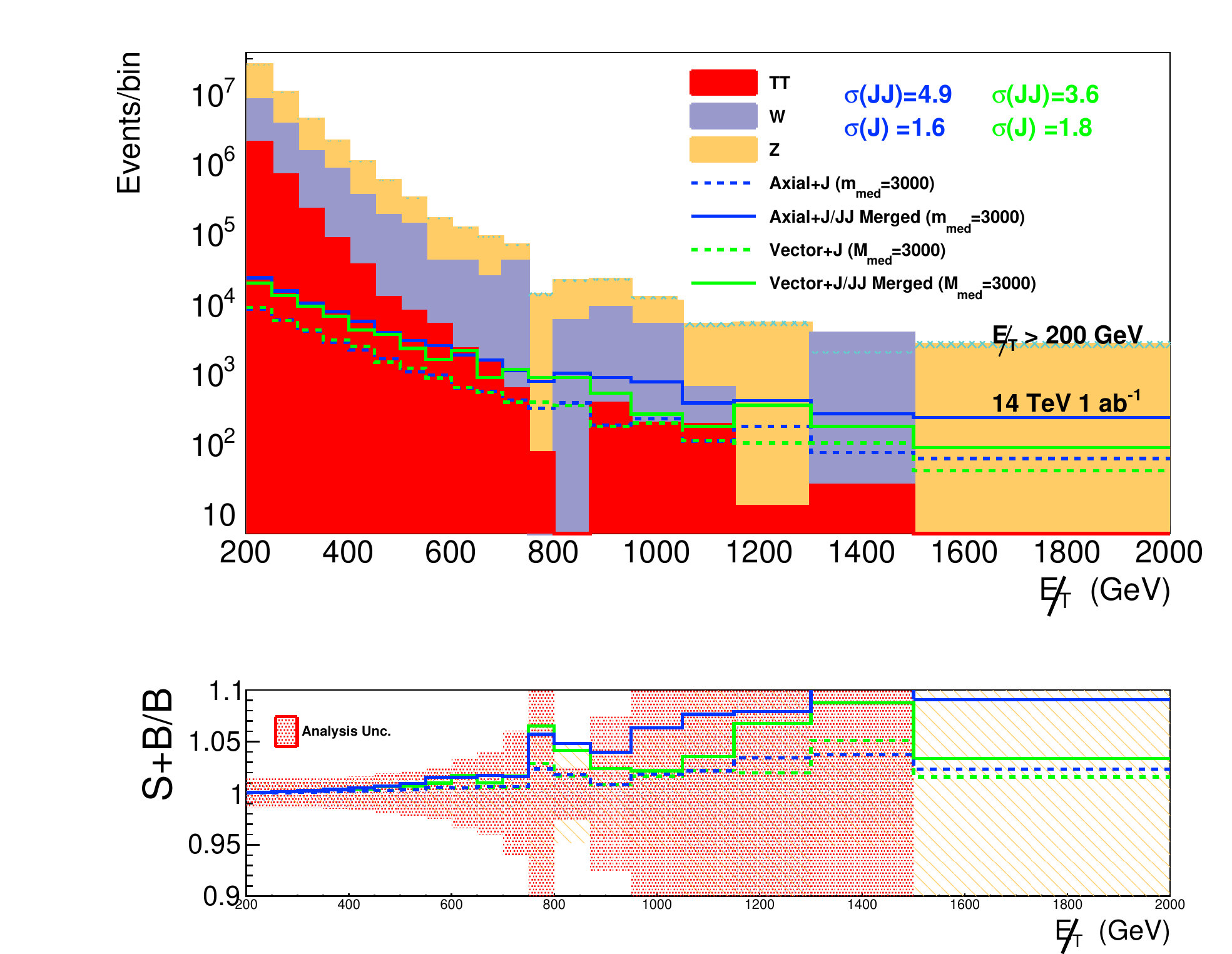}
\caption{Kinematic distributions for signal vector and axial-vector models and the SM backgrounds at $14$ TeV 
assuming 1~ab$^{-1}$ of integrated luminosity. We show four kinematic 
variables: $p_T$ of the leading jet, $p_T$ of the trailing jet, the azimuthal angle between the jets $\cos (\Delta \phi_{jj})$, 
and the missing energy \MET. Ratios of $(S+B)/B$ are shown for each observable. The red bands indicate the uncertainties on the background distributions. The accordingly color-coded numbers for sig(JJ) and sig(J) give the statistical significance to disfavour the presence of the signal using the $\mathrm{CL}_s$ method. }
\label{fig:kin14TeVaxial}
\end{figure}
\end{center}

\begin{center}
\begin{figure}[t]
\includegraphics[width=0.5\textwidth]{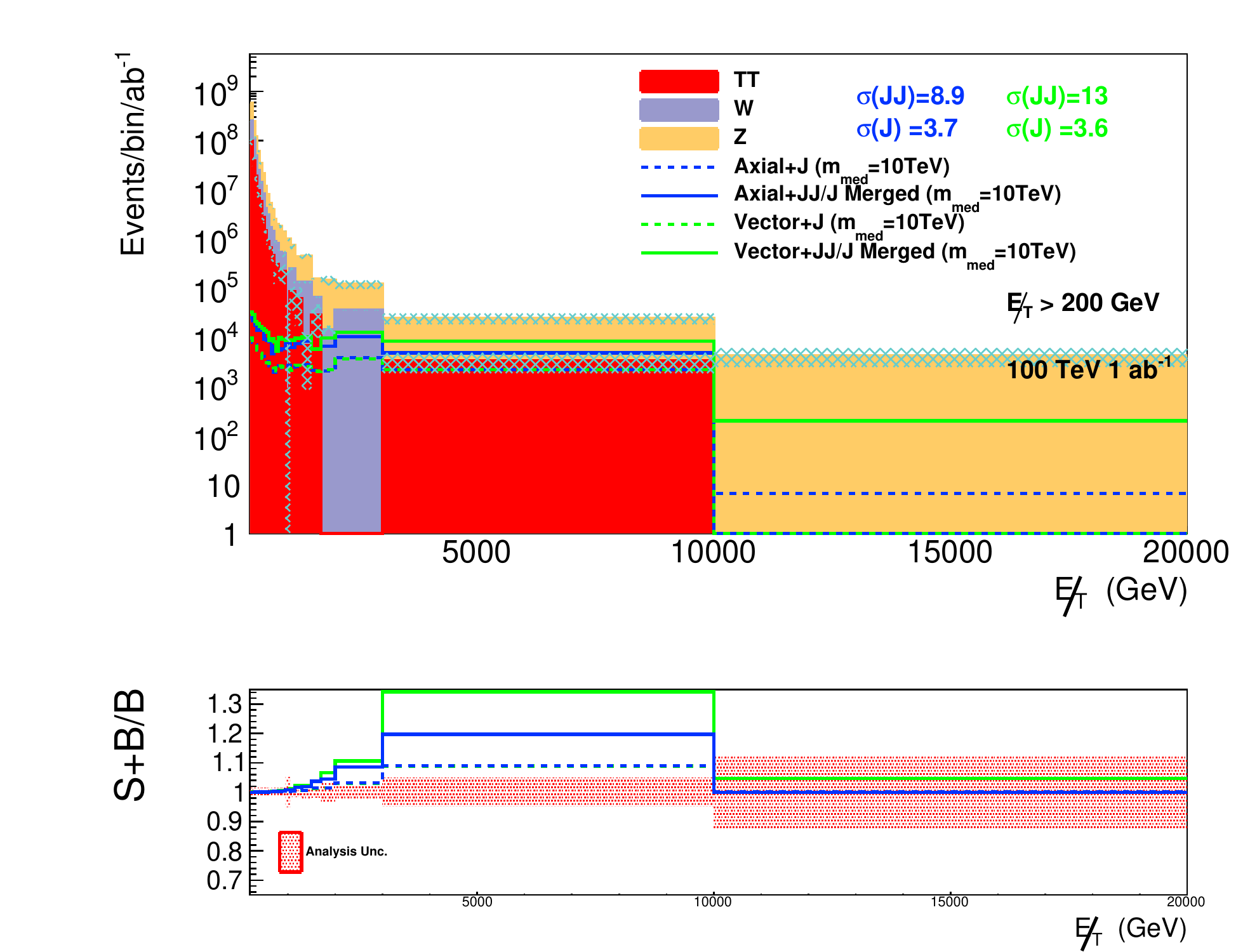} \hskip-0.5cm
\includegraphics[width=0.5\textwidth]{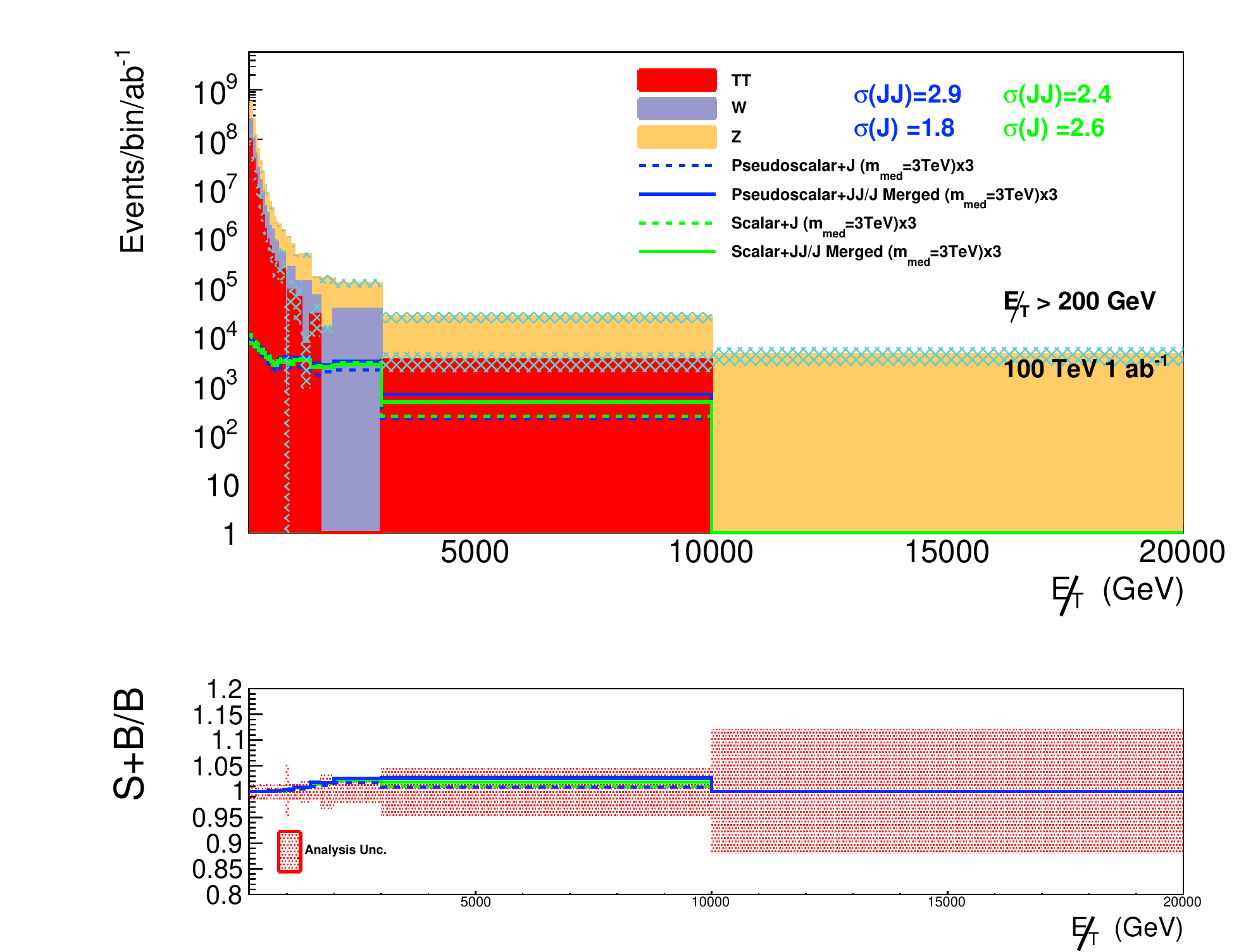}\\
\includegraphics[width=0.5\textwidth]{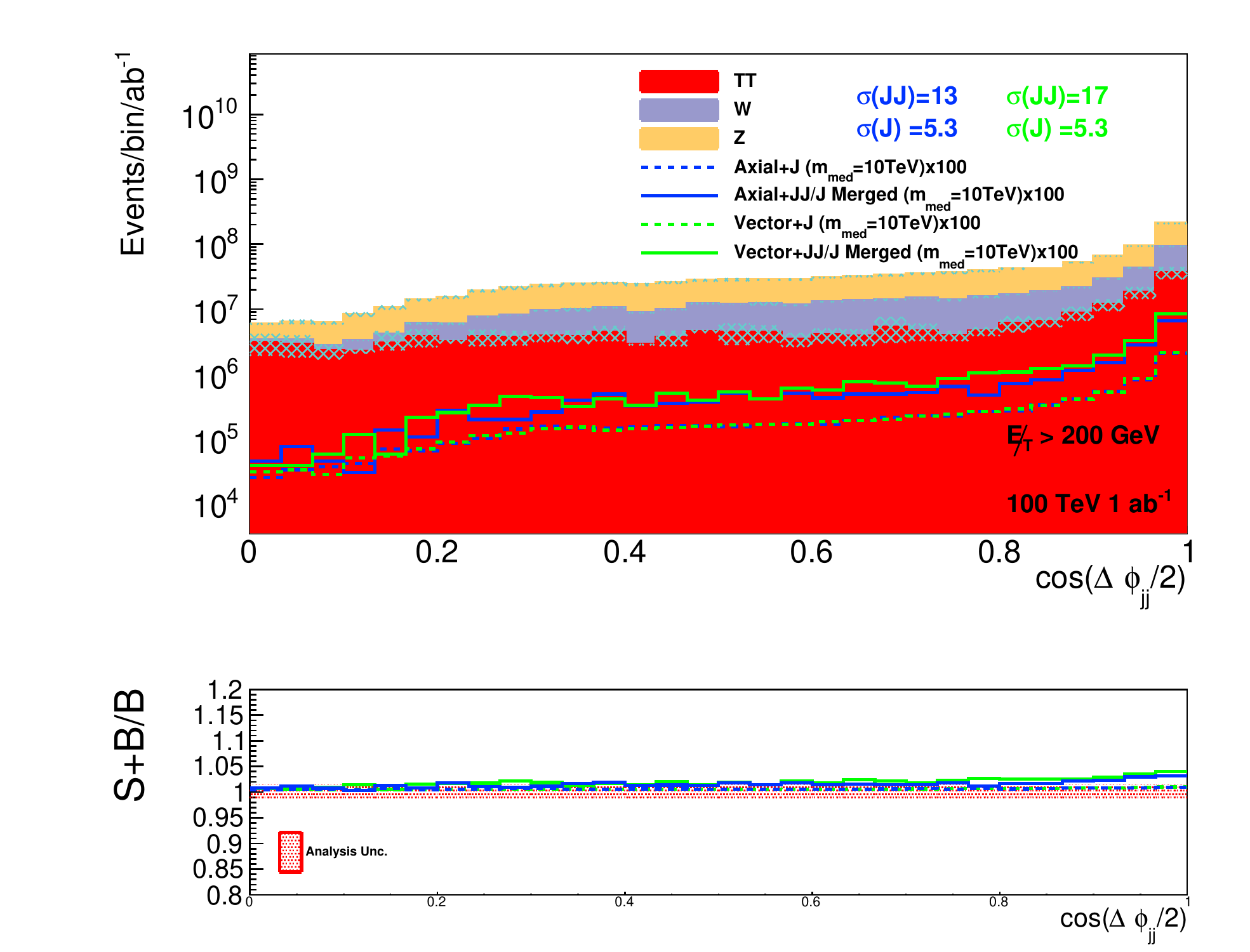} \hskip-0.5cm
\includegraphics[width=0.5\textwidth]{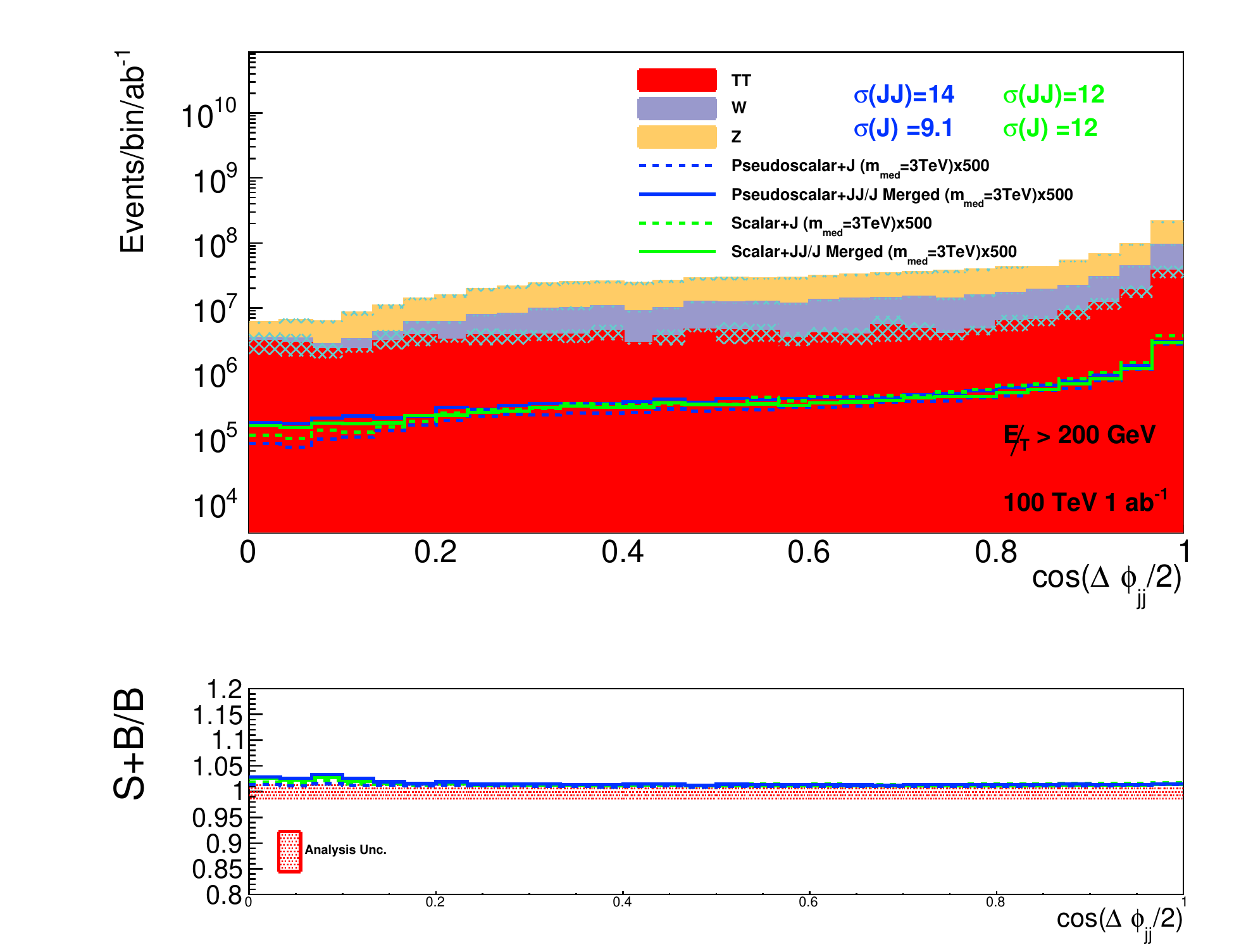}
\caption{Kinematic distributions for the two most discriminating variables \MET and $\cos (\Delta \phi_{jj})$ at $\sqrt{s}=100$ TeV.}
\label{fig:kin100TeVscal}
\end{figure}
\end{center}

Firstly, there is the obvious issue regarding tuning and PDF fits, which have been undertaken at much lower energy than those 
accessible at FCC's. Of particular importance are the extraction of the gluon PDF's, which will dominate the initial state configurations. 
Secondly, there is the accuracy of the perturbative component of the simulations. This is particularly worrisome. At FCC energies emission 
of additional radiation will result in copious jet-production around the Electroweak scale. This will require delicate handling 
with respect to matching and merging of parton shower and matrix element emissions. Since matching prescriptions typically
require scales which separate emissions into the two categories, one can easily imagine existing tools for LHC physics are not 
optimal for future FCC predictions. An additional concern relates to the simulation of Electroweak bosons, at 100 TeV, the mass of 
the $W$, $Z$ and $H$ bosons become small scales, and Sudakov logarithms associated with their emission from partons becomes 
relevant. The resolution of these issues will take many years of research and improvements to existing tools. However given the likely 
timescale of construction, and the rapid improvement in theoretical tools, none of the above issues should be regarded as significantly 
likely to negatively affect the physics program at the FCC. They should certainly be kept in mind and the resulting theoretical predictions
used in this paper should be interpreted as having a large uncertainty. 

In order to attempt to simulate events at the FCC it is therefore critical to 
include high energy jet radiation in the best manner possible. This can be achieved 
by ensuring that matched samples are used. 
Backgrounds are generated at next to leading order for 0,1,2
jets merged using  MadGraph-aMC@NLO~\cite{Alwall:2014hca}, with the
exception of $W$+jets which does not have the second jet merged. 
Diboson and single top processes are all included.  For the signal we use MadGraph for 
the Vector/Axial simplified models and a combination of
MCFM~\cite{Fox:2012ru,MCFMweb} and VBFNLO~\cite{VBFNLO} for the
production of Scalar/Pseudoscalar mediators in association with one
and two-jets. The output LHE events are then merged using the CKKW-L
interface of Pythia 8~\cite{Sjostrand:2007gs}. NNPDF3.0~\cite{Ball:2014uwa} PDF's are
used for the generation of all Monte-Carlo samples. In order to generate
the samples consistently the mediator is assumed to be on-shell for
the 1 and 2 jet merged sample. The one-jet sample is generated with
the full width and thus includes off-shell effects. Since the putative mediators scrutinized at the FCC will include heavier masses, off-shell effects will indeed play an important role in the simulation. However, given the large center of mass energy, including at least the two-jet matrix element (and ideally in the future 3-jet and beyond) is likely to be a more important requirement than including off-shell effects\footnote{We note that very recently ref~\cite{Backovic:2015soa} presented an implementation of the signal model which can incorporate one- and two-jet merged samples and off-shell effects.}.  In the following section we therefore present results both for the one-jet sample (with the full off-shell effects)  and the two-jet merged sample (which includes hard matrix element jets, but requires on-shell mediators). We note that that this is not an issue for the vector/axial mediators. In all cases we calculate the minimum width which corresponds to all open fermionic decay channels (see Eqs.~\eqref{eq:GVA}-\eqref{eq:GS} for definitions). 

In addition to the theoretical issues discussed above, we have limited knowledge of future detectors. However, here a conservative 
assumption can easily be made. We simply assume in this paper, that
the detector at the FCC is a copy of the CMS detector. One extension
to this detector is added,  the lepton
acceptance for this upgraded detector is assumed to extend up to
$|\eta| < 4.0$ and $|\eta| < 5.5$ for the 14~TeV and 100~TeV
detector respectively. This design is consistent with the plans for
the upgraded CMS and ATLAS
detectors~\cite{Butler:2020886,ATLAS:1502664}.  
The choice of detector allows us to utilize the extensive studies of the detector
performance. In reality we can expect improvements in technology which
can significantly improve on existing detectors.

To emulate the effects of a detector, the MET is smeared as a function of the mediator \pt~\cite{Khachatryan:2014gga}. The jets are smeared as a function of $\eta$ and \pt~\cite{CMSJet}. Effects from pileup are taken to account, such that the respected observed resolution follows closely the expected conditions for high luminosity running at the LHC.

\subsection{Kinematic distributions for signal and background}

In light of using multi-leg Monte Carlos and recent developments in shape based analysis for Dark Matter projections~\cite{CMS:2015jha}, the analysis selection was re-optimized from a previous cut and count analysis by considering a shape based analysis in both single and pairwise combinations of variables that take into account both the leading and trailing jets. To find sensitive variables to separate signal from backgrounds we focused on \MET, $m_{T}^{2}$, the razor variables, the scalar sum of all jets above 30 GeV ($h_{T}$), and the angle between the two leading jets $\Delta \phi_{jj}$~\cite{CMS:2015fla,CMS:2015jha,CMS:2012ova}.
Figure~\ref{fig:roc}, shows the performance of these variables when
compared with an amalgam of signal Monte Carlos. For these signal
models, a dark matter mass of 50 GeV is taken (ensuring on-shell decays for both). The range of samples consist of those with different mediator masses ranging from 1TeV to 2 TeV in 100 GeV intervals. The samples are combined by cross-section weights. We recall that the 1 jet Monte Carlo is generated with full on
and off shell width effects, whereas the 1-jet/2-jet merged sample is
generated explicitly on-shell. When considering the single variable performance, we find \MET is the most discriminating variable. As a result, combining these variables into pairwise combinations using a boosted decision tree we find that the most sensitive combination is \MET and $\Delta\phi_{jj}$ between the two leading jets.  
This can be readily seen in the left plot in Fig.~\ref{fig:roc} where the $\Delta\phi_{jj} +$\MET combination is shown as the solid line in blue.
In the lower-left quadrant of the plot the reconstruction efficiency of the signal $\epsilon_{\rm sig}$ significantly exceeds that of the background,
for example for $\epsilon_{\rm sig} \simeq 0.5$ we have $\epsilon_{\rm bkg} \simeq 0.1$ on the blue contour in Fig.~\ref{fig:roc}.
A comparison with higher dimensional sets of variables in the right
panel shows even further improvement, particularly when the second jet merged MC is used.
However, for this further optimization to manifest
itself in the analysis requires
precise knowledge of the kinematic discrimination of \MET in the far
tails of the distribution. Additionally we note that the larger differences
between the one-jet and two-jet signal MC, can be indicative of greater 
sensitivity to theoretical modeling. Thus, we do not extend these studies beyond
their current projections. Our final extractions are then performed by
fitting the \MET distribution in two bins of $\Delta\phi_{jj}$
($\Delta\phi_{jj} < 1.1$ and $\Delta\phi_{jj} > 1.1$). The binning is
optimized such that the the full spectrum is covered with MC
samples. Additionally a selection of  $\min ( \Delta
\phi_{\slashed{E}_T,j_i}) \geq 0.5$, where $i$ runs over all jets is applied to minimize the
impact of QCD multi-jet events where one or several of the jets are
miss-measured. As a result of this selection being applied, this
background is neglected throughout the course of these studies.  
Further improvements could be obtained by using finer bins for the tails of the distribution\footnote{While we will not include this approach in our analysis we point out that this could be a viable way to optimize the sensitivity of the search.}.

The signal extraction is performed with a full shape analysis in which the
shape for the dominant backgrounds are taken from control
regions. For the $Z\rightarrow \nu\bar{\nu}$ background, the $Z\rightarrow\mu^+\mu^-$
control region is used as a background. For the $W\rightarrow \ell \nu$,
top and diboson backgrounds, we use the single lepton control region. For each of these control regions the full statistical
uncertainty on the shape is propagated per bin on each of the
backgrounds with an additional one percent uncertainty uncorrelated
per bin to account for additional modelling uncertainties. For all
but the tail bins of the shape uncertainties on the \MET spectrum are
roughly 1\% with the dominant uncertainty resulting from the
additional one percent modelling uncertainty. Finally, the signal is
profiled using the standard limit extraction ($\mathrm{CL}_{s}$) \cite{Read:2002hq,Cowan:2010js}. Additional
nuisances are placed on the background normalization for lepton
efficiencies and luminosity. These are constrained to very small
values due to the large dataset in the signal region and do not affect
the limit sensitivity. 

Kinematic distributions for our simplified models of dark sectors alongside the main SM backgrounds 
are shown in Figs.~\ref{fig:kin14TeVscal}, \ref{fig:kin14TeVaxial} and \ref{fig:kin100TeVscal}. The distributions are shown as functions of four kinematic variables, $p_T$ of the leading jet, $p_T$ of the trailing jet, the azimuthal angle between the jets $\cos (\Delta \phi_{jj})$, and missing energy \MET.

In Figs.~\ref{fig:kin14TeVscal} and \ref{fig:kin14TeVaxial} we show
shapes of kinematic distributions corresponding to the (pseudo-)scalar
and (axial-)vector case respectively, with the signals and backgrounds
computed at $14$~TeV centre-of-mass energy and assuming 1~ab$^{-1}$
integrated luminosity. The only event selection cuts imposed for the
distributions in Figs.~\ref{fig:kin14TeVscal}-\ref{fig:kin100TeVscal}
are $\slashed{E}_T \geq 200$ GeV and $\min ( \Delta
\phi_{\slashed{E}_T,j_i}) \geq 0.5$, where $i$ runs over all jets in
each event. 

We find that for the event selection cuts imposed, the distributions for samples using multi-jet merging are similar to the unmerged samples. In the gluon-induced scalar and pseudoscalar cases the unmerged samples result in a harder $p_T$ distribution, particularly for the trailing jet. On the other hand, the $\cos (\Delta \phi_{jj})$ distribution changes more rapidly when including multi-jet merging.  
In Figure \ref{fig:kin100TeVscal} we show the kinematic distributions for the two most discriminating variables, which, as was pointed
out above, are the missing transverse energy \MET and the azimuthal angle between the jets $\Delta \phi_{jj}$ at $\sqrt{s} = 100$~TeV. The 
two plots on the left show the vector and axial-vector case while the two plots on the right show the scalar and pseudoscalar signals.


\subsection{Direct and Indirect Detection Limits}
\label{sec:limits}

In Sec.~\ref{sec:results} we will compare collider limits with limits from direct and indirect detection experiments. For convenience we recall the formulas for direct and indirect detection limits, in terms of the simplified models of Eq.~(\ref{eq:LS})-(\ref{eq:LA}).

We use 
\begin{equation}
\sigma_{\mathrm{\chi p}}^V = \frac{9}{\pi} \frac{g_{\rm DM}^2 g_{\rm SM}^2 \rho^2}{m^4_{\mathrm{MED}}}
\label{eq:16}
\end{equation}
and 
\begin{equation}
\sigma_{\mathrm{\chi p}}^A = \frac{3}{\pi}\frac{g_{\rm DM}^2 g_{\rm SM}^2 a^2 \rho^2}{m^4_{\mathrm{MED}}},
\label{eq:18}
\end{equation}
with $a \simeq 0.43$ \cite{Cheng:2012qr, Buchmueller:2013dya} and the reduced mass $\rho =m_{{\rm DM}} m_{p}/(m_{\rm DM} + m_{p})$, for the cross section of a dark matter particle scattering spin-independently (vector mediator) or spin-dependently (axial-vector mediator) from a proton.

The cross section for a Dark Matter particle scattering from a nuclei via a scalar mediator of Eq.~(\ref{eq:LS}) is given by 
\cite{Kurylov:2003ra,Hisano:2010ct, Cheung:2013pfa}
\begin{equation}
\sigma_{\mathrm{\chi p}}^S =  \frac{\rho^2}{\pi}  \left | \sum_{q=u,d,s} f^{p}_q \,\frac{m_p}{m_q}\left( \frac{g_{DM} g_q y_q}
{m^2_{\mathrm{MED}}} \right )    
+ \frac{2}{27} f_{\mathrm{TG}} \,\sum_{q=c,b,t} \frac{m_p}{m_q}\left( \frac{g_{DM} g_q y_q}{m^2_{\mathrm{MED}}} \right ) \right |^2,
\label{eq:17}
\end{equation}
where $f^{p}_u= 0.019$, $f^{p}_d=  0.045$, $f^{p}_s= 0.043$ and 
$f_\mathrm{TG}\simeq 1-\sum_{q=u,d,s}f^{n}_{q} $ \cite{Hoferichter:2015dsa,Crivellin:2013ipa}
and $m_p$ is the proton mass.

When comparing the expected sensitivity for the LHC and FCC for DM searches to those of Direct Detection it is interesting to compare to 
expected impact of the neutrino wall~\cite{Cushman:2013zza,Buchmueller:2014yoa}. The neutrino wall is a result of the background of cosmic neutrinos. We take their interaction cross section to be indicative for the ultimate reach of DD experiments. Therefore in the following section we will present the equivalent limit which may be 
obtained from DD experiments with the currently hypothesized values of the neutrino wall ~\cite{Cushman:2013zza,Buchmueller:2014yoa}.  
For a pseudo-scalar mediator, taking existing limits into account \cite{Ackermann:2011wa,Abdo:2010ex}, indirect detection experiments can result in stronger limits than direct detection experiments \cite{Zheng:2010js, Boehm:2014hva}. For the simplified model of Eq.~(\ref{eq:LP}), using the velocity-averaged DM annihilation cross section into $\bar{b}b$,
\begin{equation}
\left < \sigma v \right >_{\bar{b}b}^P = \frac{N_C}{2 \pi} \frac{(y_b g_{b})^2 g_{DM}^2  \, m_{\rm DM}^2 } {(m_{\rm MED}^2 - 4 m_{\rm DM}^2)^2 + m_{\rm MED}^2 \Gamma_{\rm MED}^2} \sqrt{1 - \frac{m_b^2}{m^2_{DM}}},
\label{eq:19}
\end{equation}
which allows us to derive a limit on the parameters in the $\bar{b}b$ channel \cite{Ackermann:2011wa}. 

\section{Results} 
\label{sec:results}

Results are obtained scanning over a spectrum of signal models at 14 TeV
and 100 TeV. A predicted luminosity of 1~ab$^{-1}$ is used for both
analyses, so the sensitivity can be compared directly.  We note that this 
amount of integrated luminosity is a rather modest amount compared to what is likely 
to be collected at a future collider, whereas for the LHC it corresponds to a significant 
fraction of the data set that will be obtained.
As a result the results we present here should be interpreted as those which can be obtained over the lifetime 
of the LHC, and for a shorter run with the FCC.  

\begin{center}
\begin{figure}[h]
\includegraphics[width=0.51\textwidth]{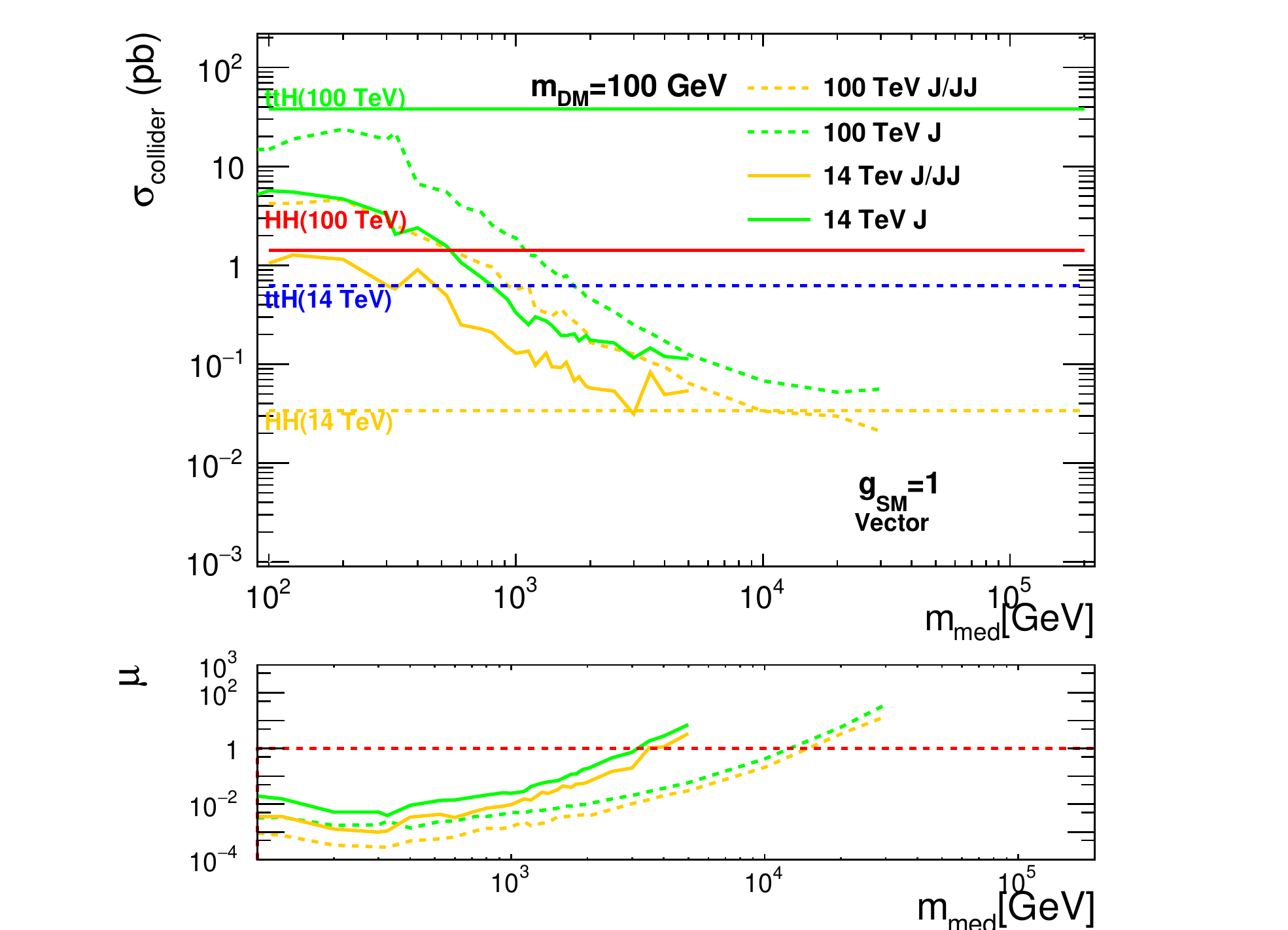} \hskip-0.9cm
\includegraphics[width=0.51\textwidth]{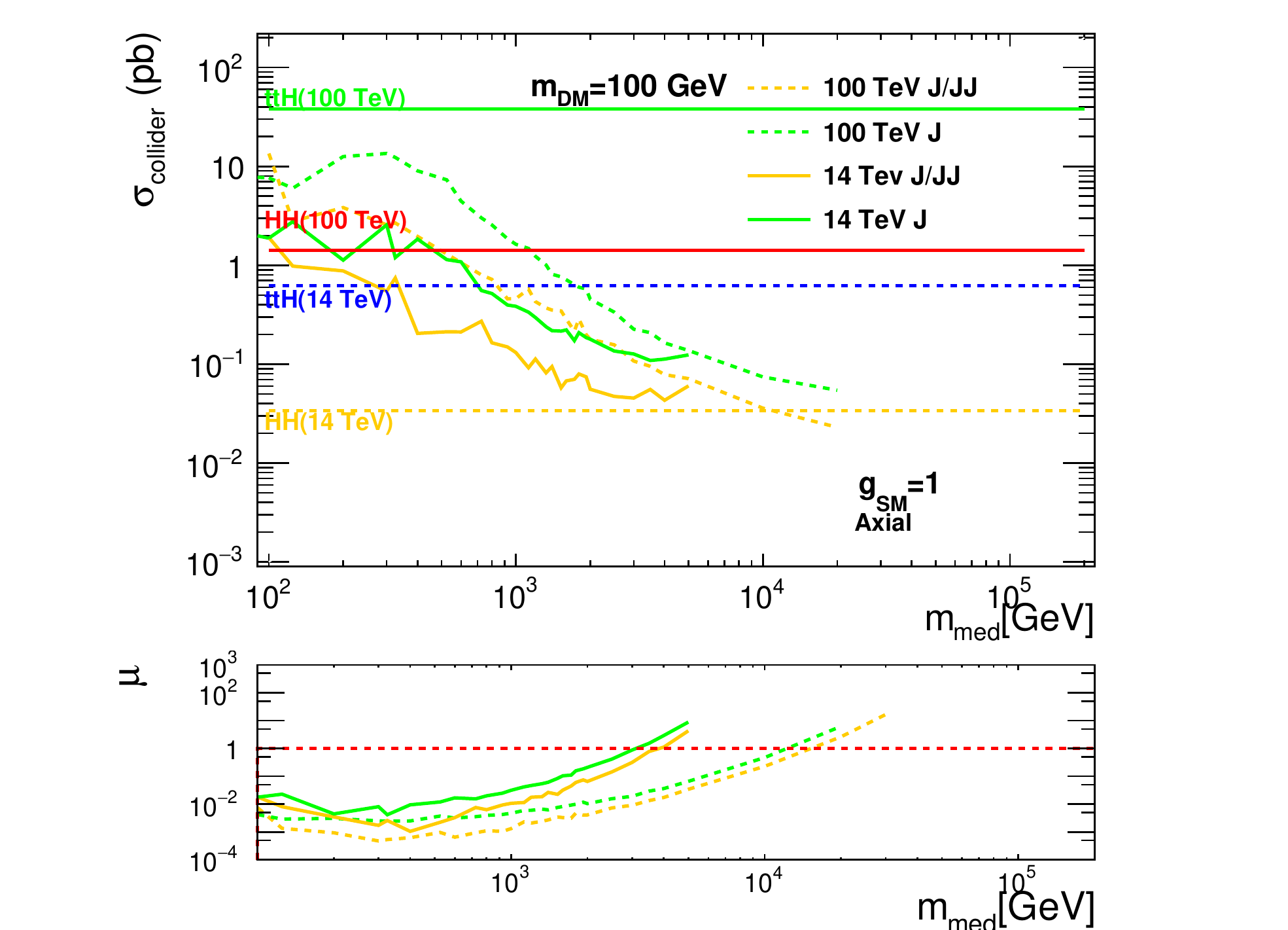}\\
\includegraphics[width=0.51\textwidth]{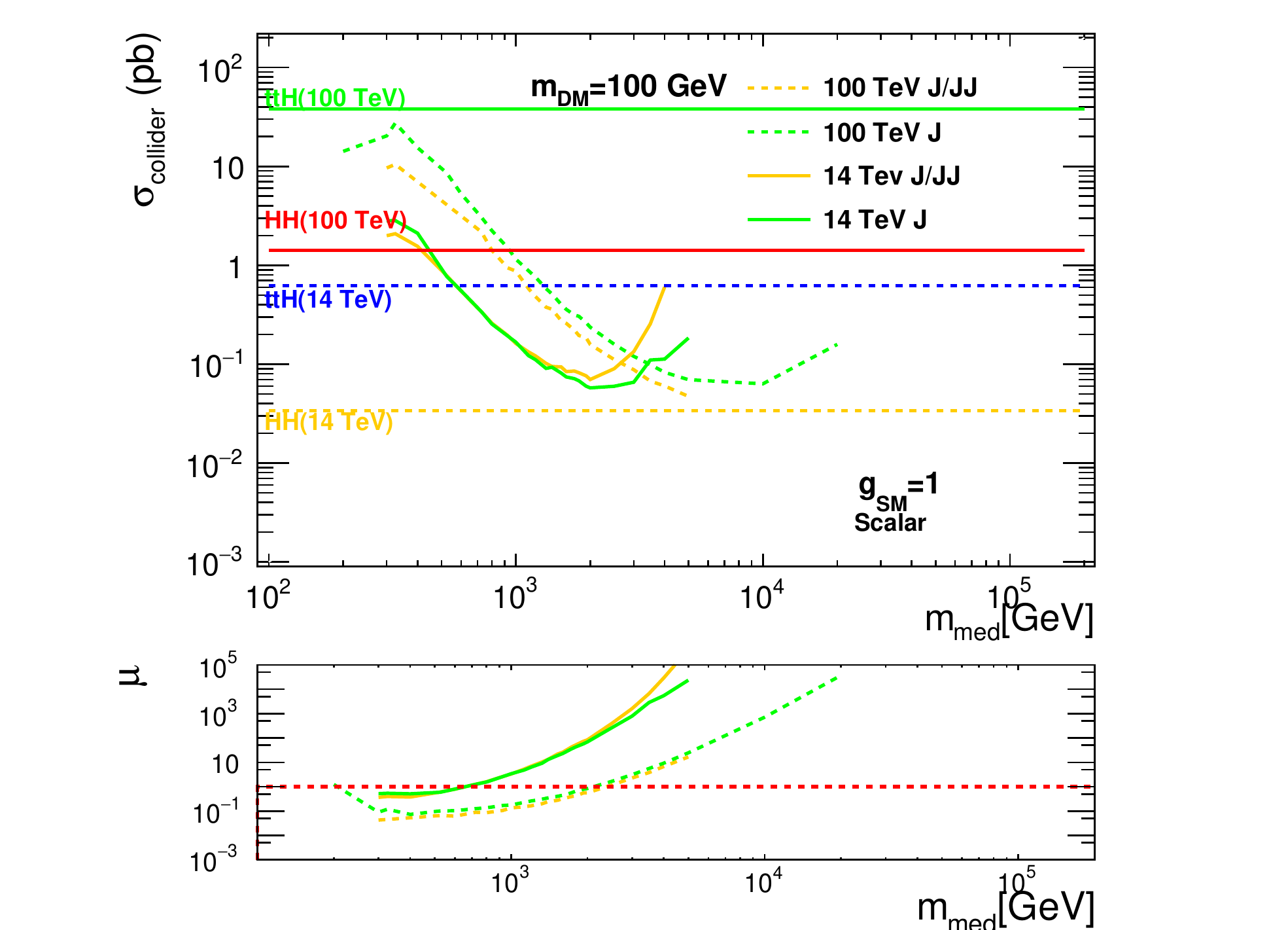} \hskip-0.9cm
\includegraphics[width=0.51\textwidth]{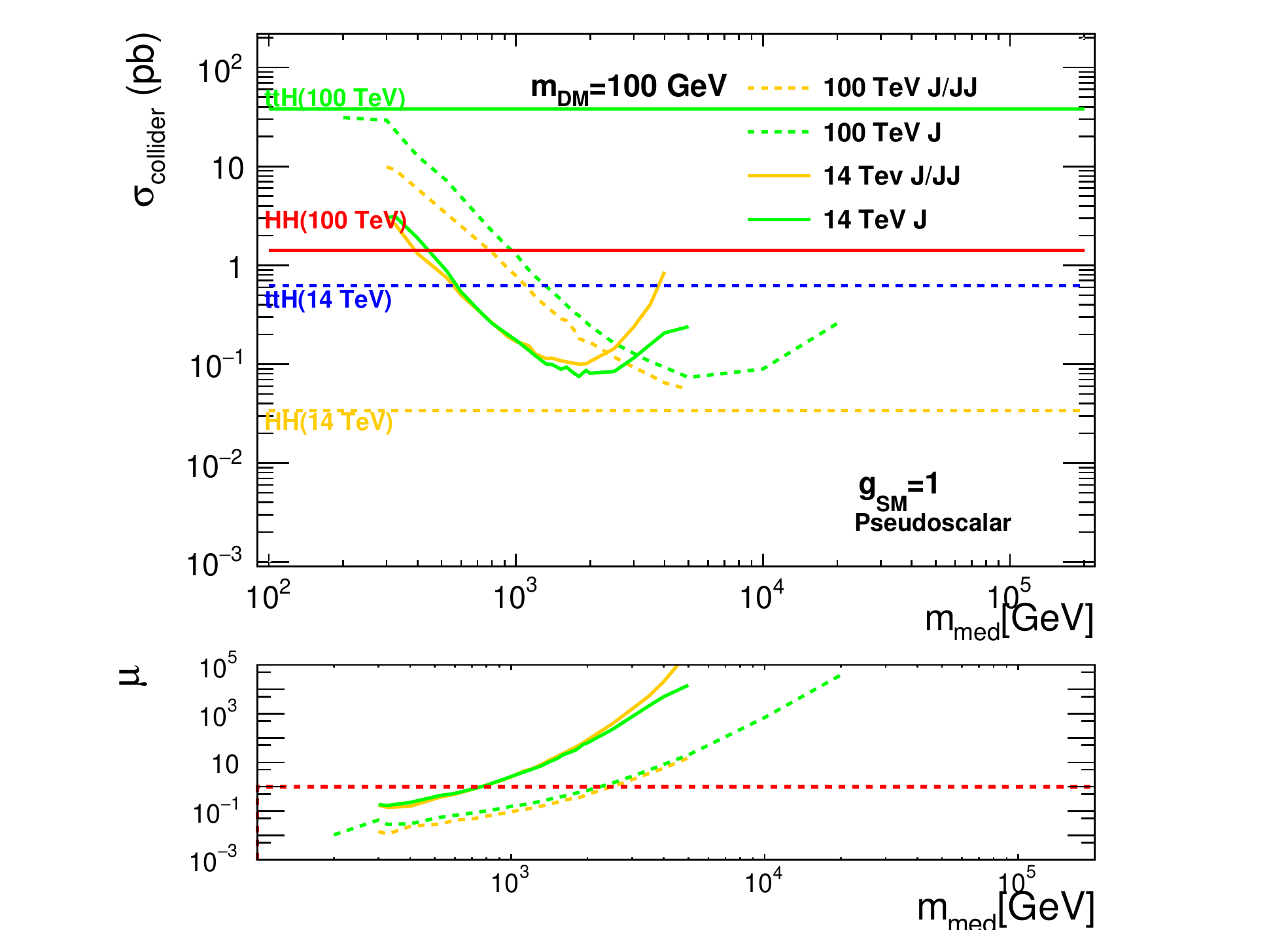}
\caption{Cross section exclusion limits as a function of mediator mass for a fixed dark
matter mass   at a given coupling. We show results for vector (upper left panel), axial-vector (upper right panel), scalar (lower left panel) and pseudoscalar (lower right panel)}
\label{fig:exclCrossS}
\end{figure}
\end{center}

We begin by studying constraints on total cross sections which can be obtained using our analysis. 
Figure~\ref{fig:exclCrossS} presents the total cross section which the analysis excludes for each of the four mediator types defined in Eqs.~\eqref{eq:LS}-\eqref{eq:LA}.
We define our cross sections by setting $g_{\mathrm{DM}} = g_{\mathrm{SM}} = 1$ and select the mediator mass as indicated in the legend of each figure respectively. 
As an illustrative example we have chosen a relatively small characteristic value of 100 GeV, although the results 
obtained for other kinematically accessible values of dark matter mass were found to be similar.  
The kinematics of the process are then completely specified once the couplings $g_{\rm DM}$ and $g_{\rm SM}$  are set, since this fixes the minimal width of the mediator
\cite{Harris:2014hga}. The excluded cross section is then related to the predicted cross section as follows, 
\begin{equation}
\label{eq:mu}
\sigma = \mu ~ \sigma(g_{\rm DM} = 1, g_{\rm SM}=1, m_{\rm MED}),
\end{equation} 
With the kinematics of the model fixed we set a limit on $\mu$ defined above using the $\mathrm{CL}_s$-method, again  assuming $1~\mathrm{ab}^{-1}$ of data. Values with $\mu<1$  indicate the excluded couplings and  width are smaller than the tested model, and the point is then excluded.

In Fig.~\ref{fig:exclCrossS} we also distinguish between the mono-jet (shown in green) and the multi-jet-based analyses (shown in yellow).
It can be seen that the new multi-jet-based analysis is more powerful and provides a considerable improvement 
at 14 and at 100 TeV.
It readily follows from Fig.~\ref{fig:exclCrossS} that the new multi-leg analysis provides a considerable improvement at 14 TeV relative 
to the results of our earlier work  \cite{Harris:2014hga}. 
At 100 TeV the exclusion limits in Fig.~\ref{fig:exclCrossS} are completely new, and we also point out that the $\mu$-factor
remains $\lesssim 1$ well into the TeV range of mediator masses even for scalar and pseudo-scalar mediators.
The impact of including the additional matrix elements in the signal simulation is, as predicted, much greater for the FCC, which 
allows for copious production of light degrees of freedom. 
For the case of scalar and pseudo-scalar mediators at 14 TeV there is a cross-over for mediators heavier
than $\simeq 1$ TeV, which is absent at 100 TeV. This corresponds to exactly the regions of phase space in which the off-shell 
effects dominate. The one-jet sample has access to the significant cross section which arises from the tails of the Breit-Wigner distribution, whereas the 
multileg sample does not. This region therefore has large theory errors using the multi-leg sample. However, we note that the region of phase space 
for which the multi-leg sample breaks down is far from the values of $\mu =1$, so this region of phase space is of limited importance in regards to setting 
limits on model parameters. 
Finally we note that Fig.~\ref{fig:exclCrossS} also includes cross sections for interesting SM predictions which the FCC and Run II of the LHC will investigate.  
We present the cross sections for $t\overline{t}H$ and $HH$ and show their relative size compared to our DM predictions. 

\begin{center}
\begin{figure}[h]
\includegraphics[width=0.45\textwidth]{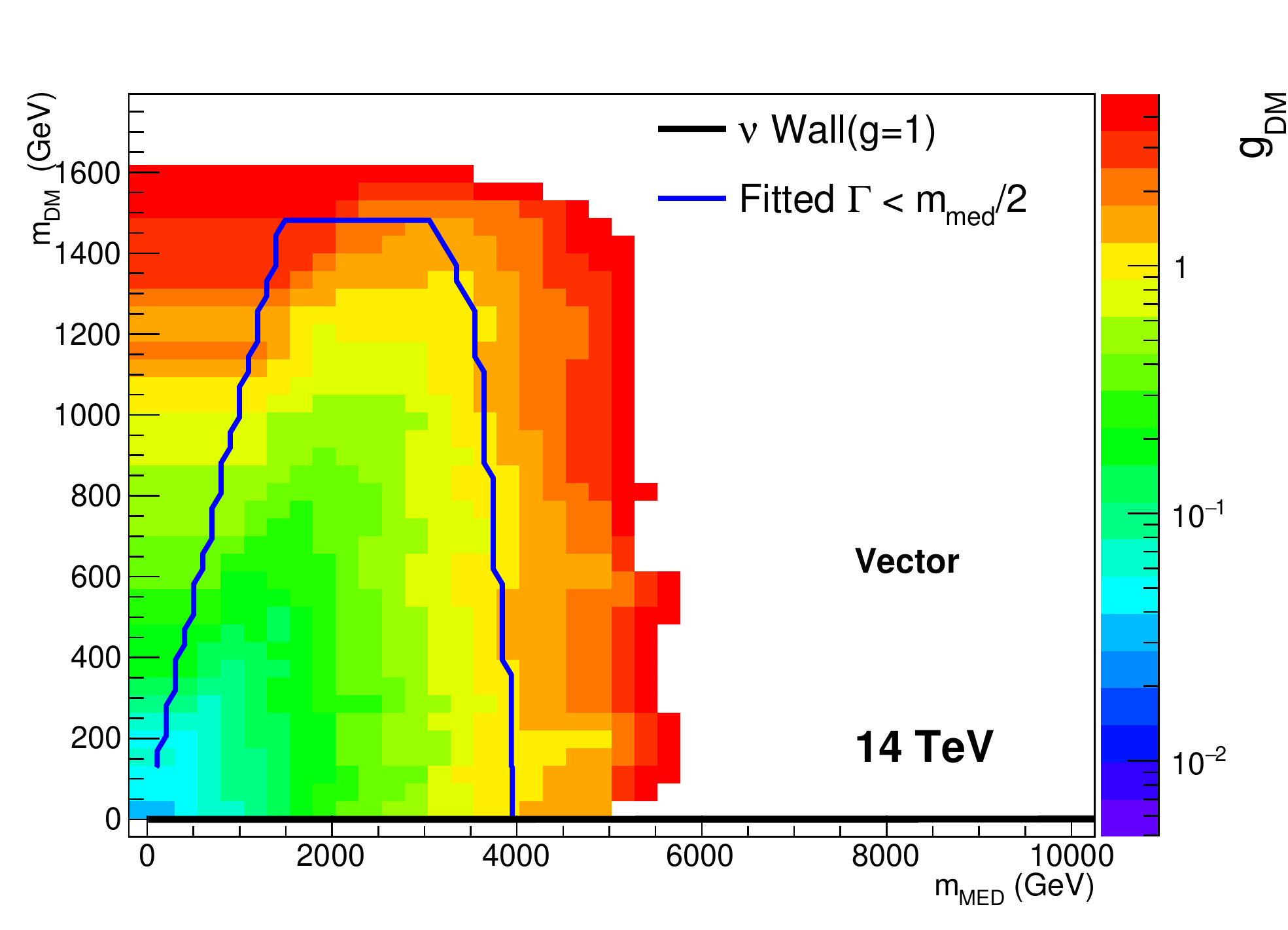}
\includegraphics[width=0.45\textwidth]{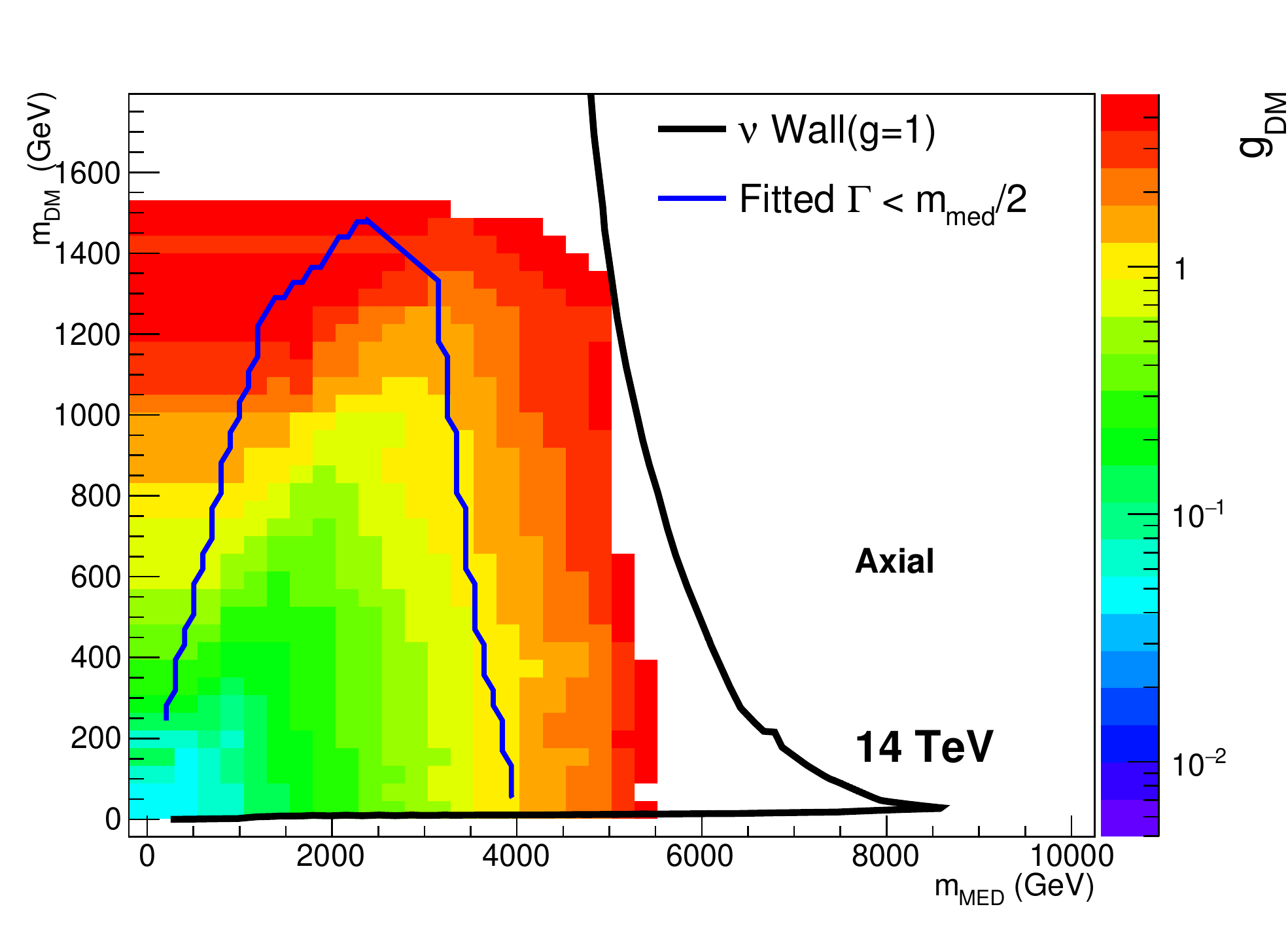}\\
\includegraphics[width=0.45\textwidth]{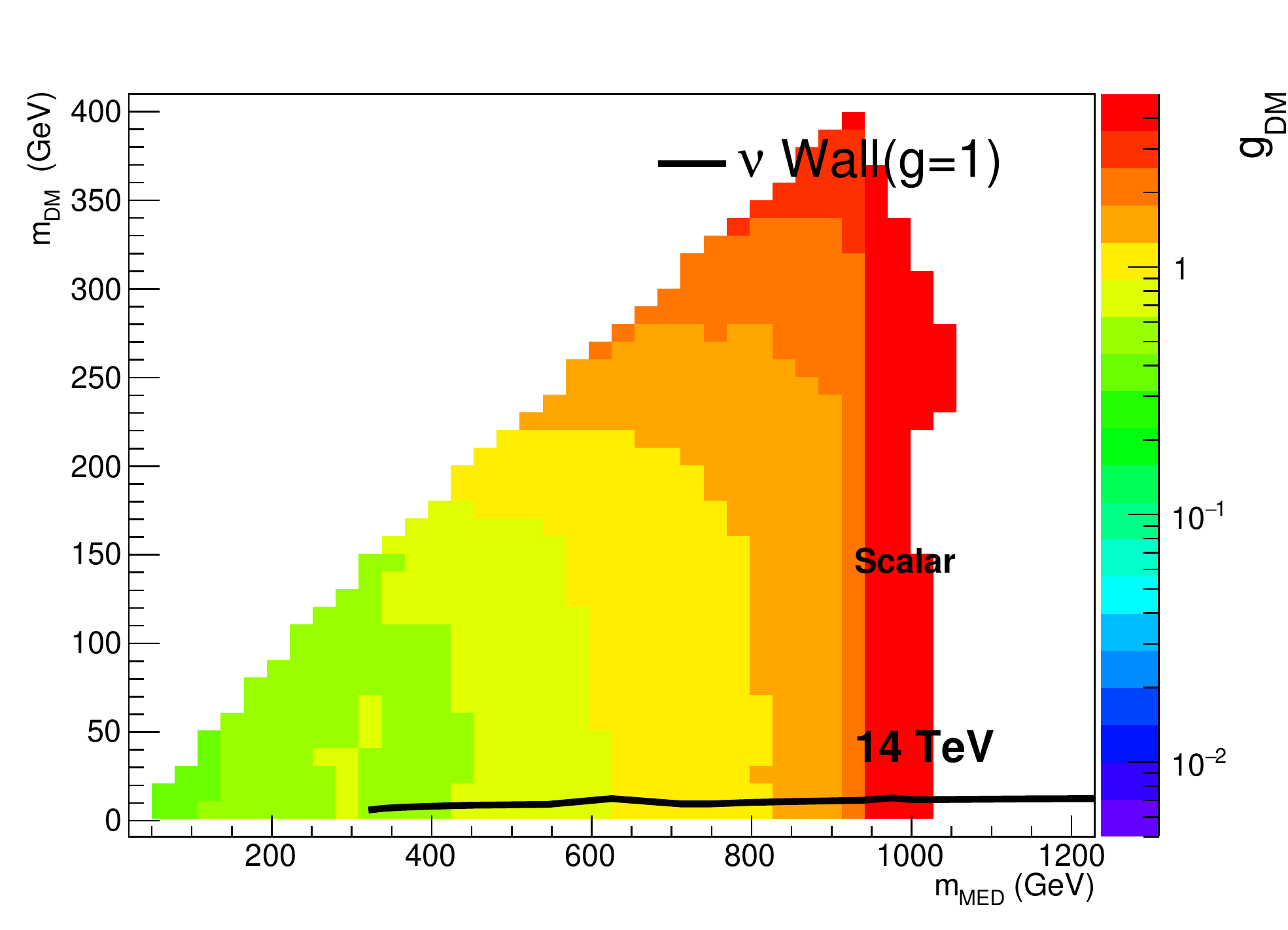}
\includegraphics[width=0.45\textwidth]{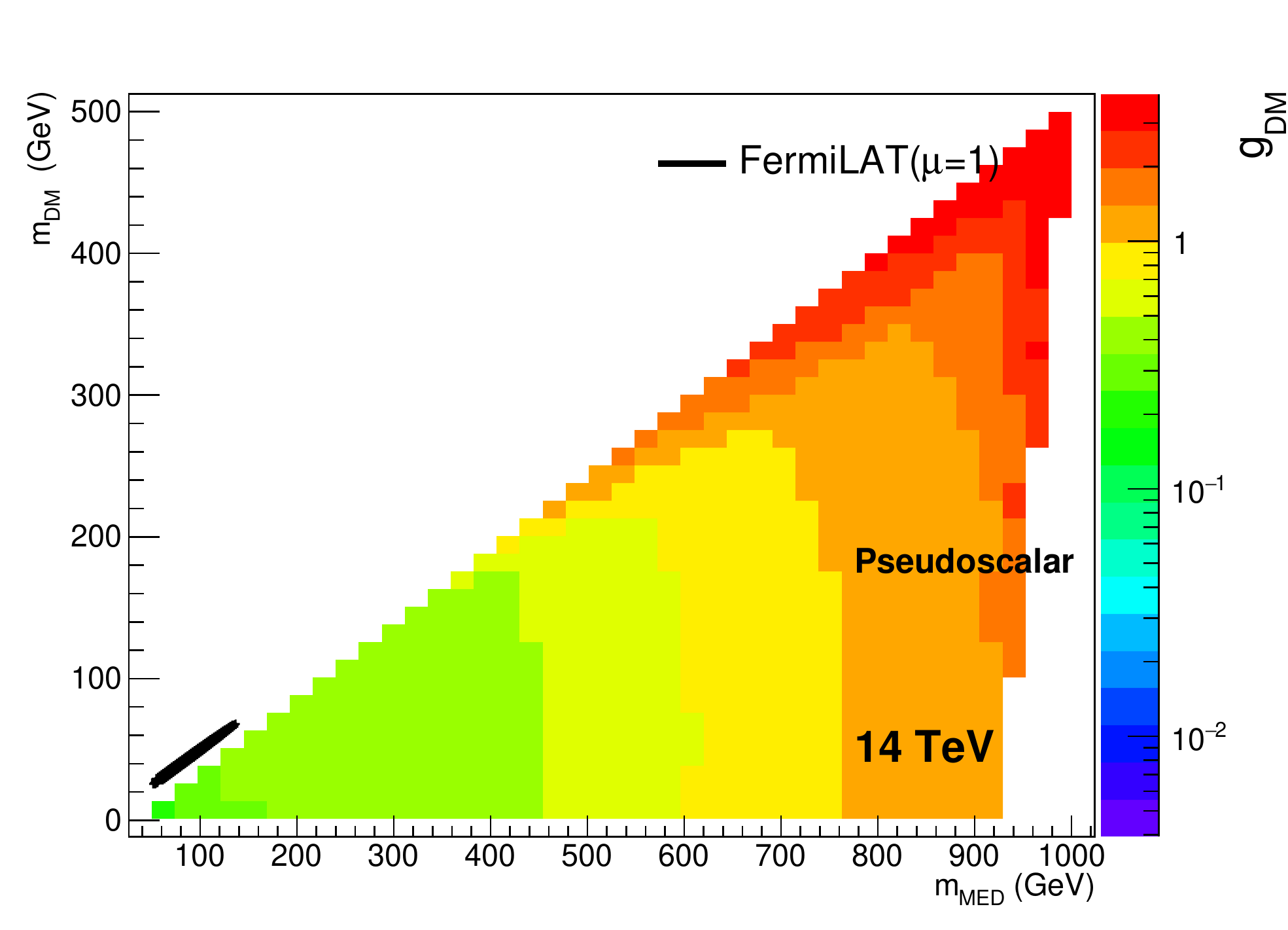}
\caption{Exclusion limits on mediator mass versus dark matter mass for given couplings based on multijet measurements
at 14 TeV LHC. We show results for vector (upper left panel), axial-vector (upper right panel), scalar (lower left panel) and pseudo-scalar (lower right panel) mediators. Plots for vector and axial-vector mediators also show the regions inside which 
$\Gamma_{\rm MED,\ min} < m_{\rm MED}$ so that the particle physics interpretation of (axial)-vector mediators applies. 
Scalar and pseudo-scalar mediators widths we use are sufficiently narrow.}
\label{fig:excl14TeV_Mmed_mDM}
\end{figure}
\end{center}

\begin{center}
\begin{figure}[h]
\includegraphics[width=0.45\textwidth]{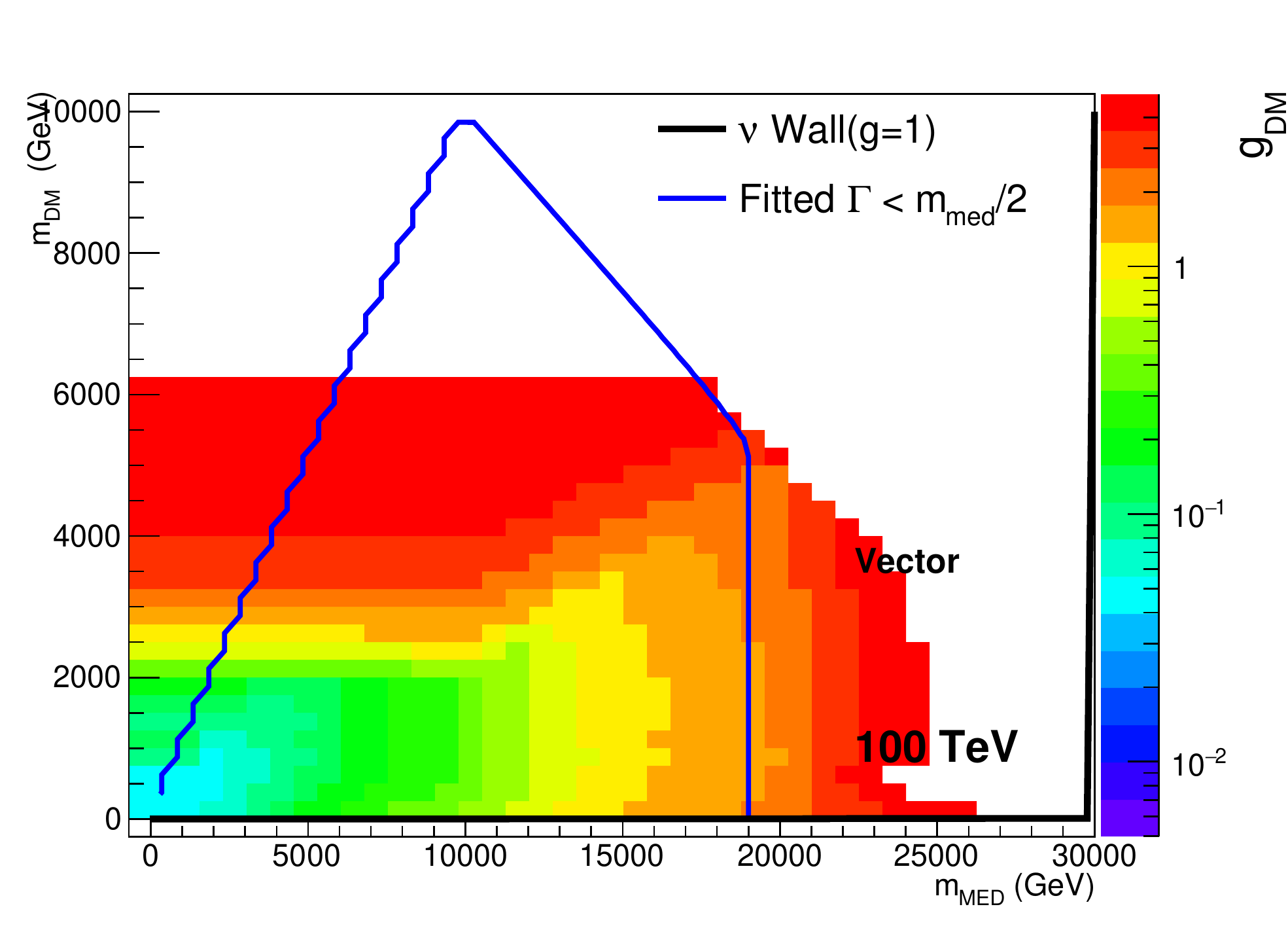}
\includegraphics[width=0.45\textwidth]{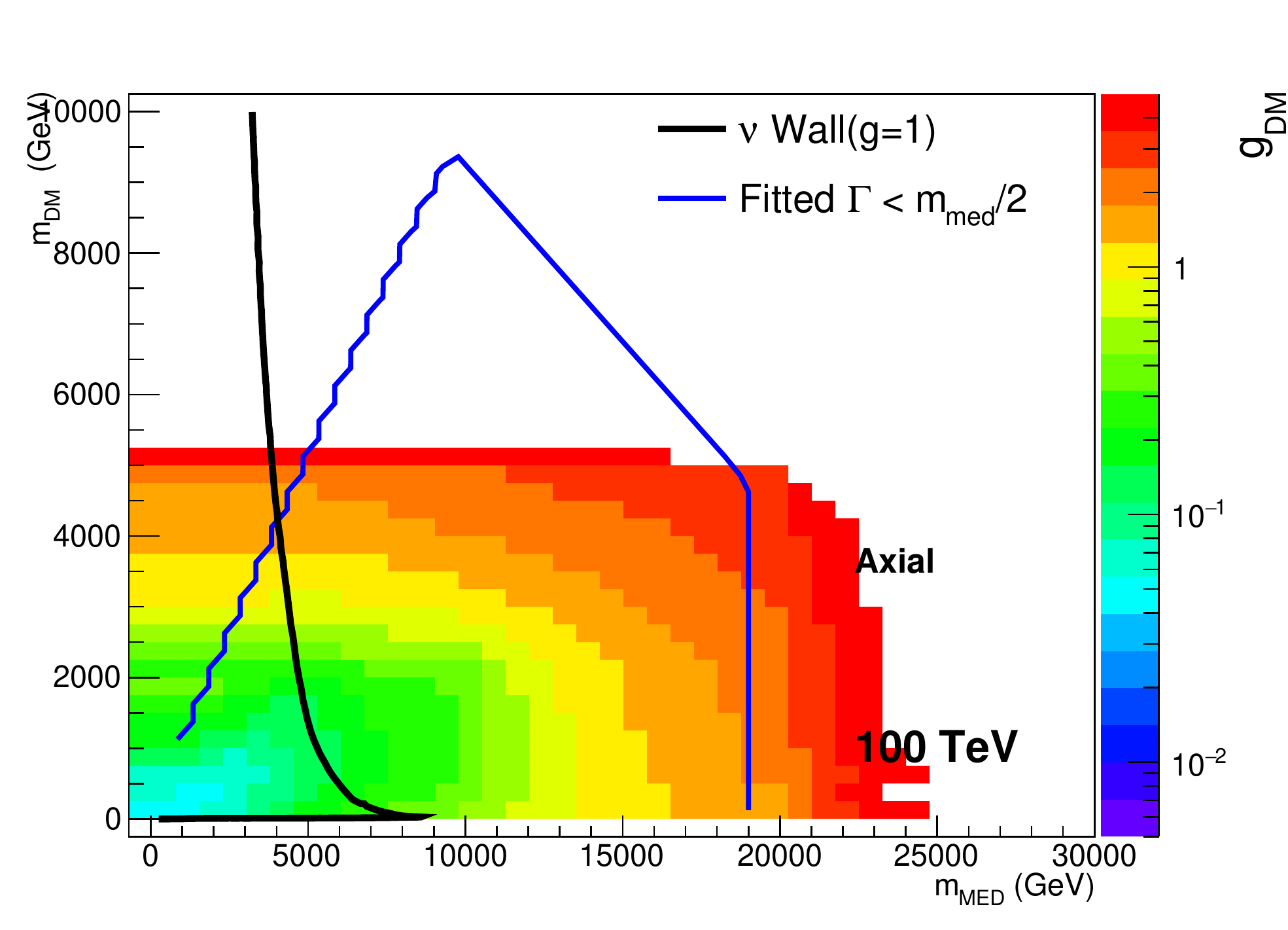}\\
\includegraphics[width=0.45\textwidth]{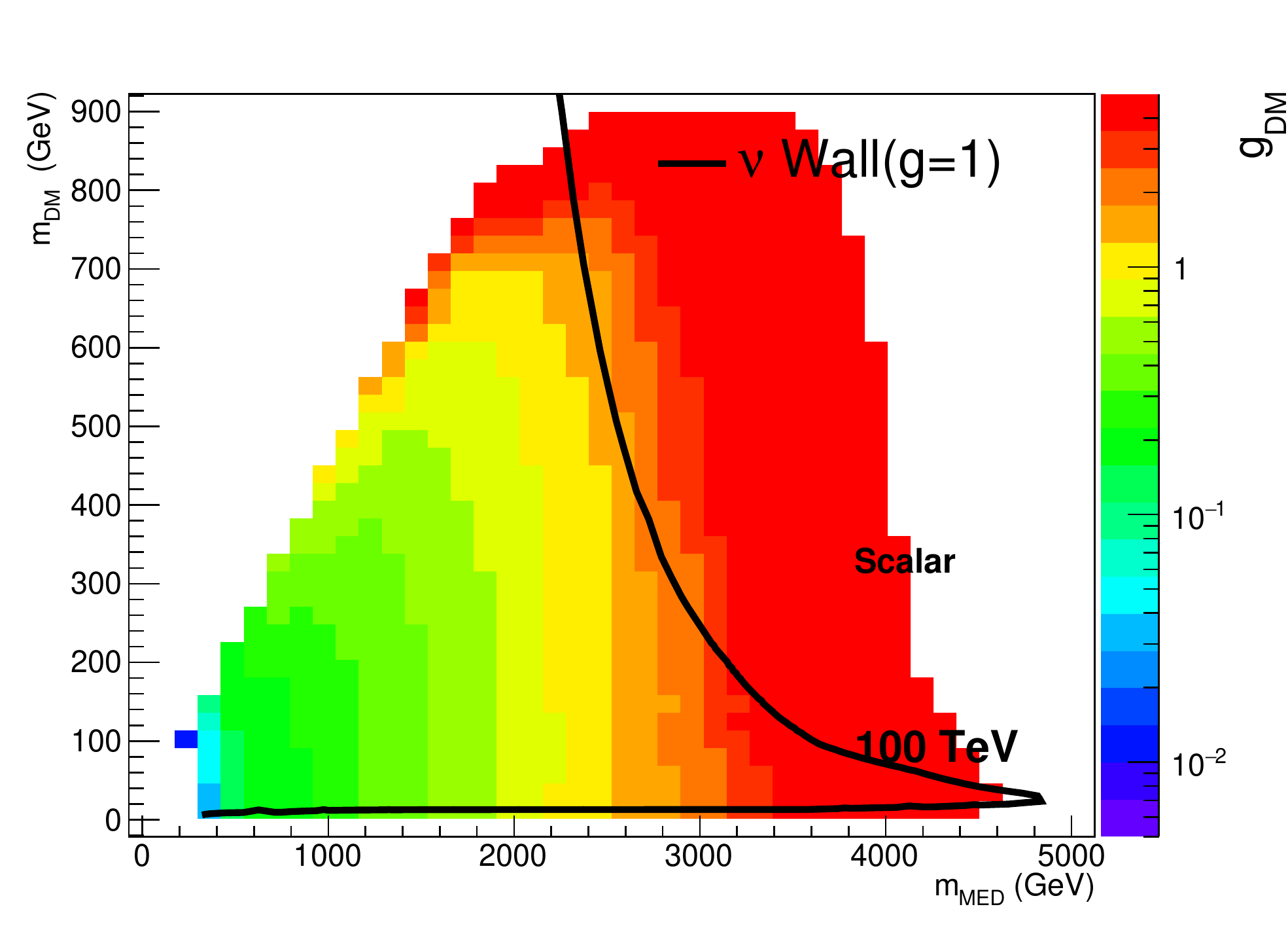}
\includegraphics[width=0.45\textwidth]{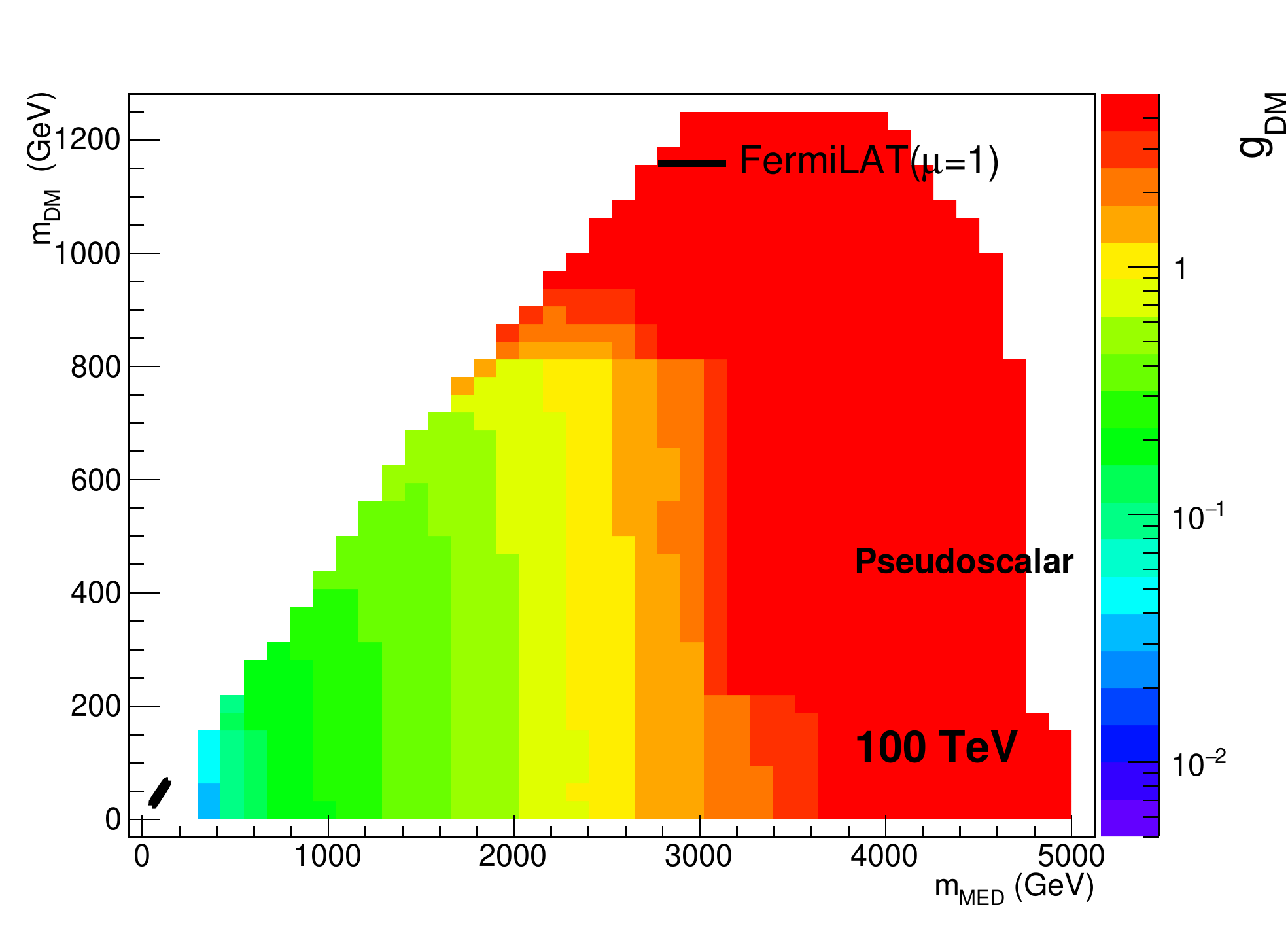}
\caption{100 TeV exclusion limits on mediator mass versus dark matter mass for a given coupling g based on multijet measurements. We show results for vector (upper left panel), axial-vector (upper right panel), scalar (lower left panel) and pseudoscalar (lower right panel).
Vector and axial-vector mediators plots indicate the regions where the minimal mediator widths do not exceed 
$m_{\rm MED}/2$, thus reducing the available parameter space reach to $m_{\rm DM}\lesssim m_{\rm MED}/2$ and cutting off 
large values of the couplings. For (pseudo)-scalar mediators the width we use is always sufficiently narrow. }
\label{fig:excl100TeV_Mmed_mDM}
\end{figure}
\end{center}

\begin{center}
\begin{figure}[h]
\includegraphics[width=0.5\textwidth]{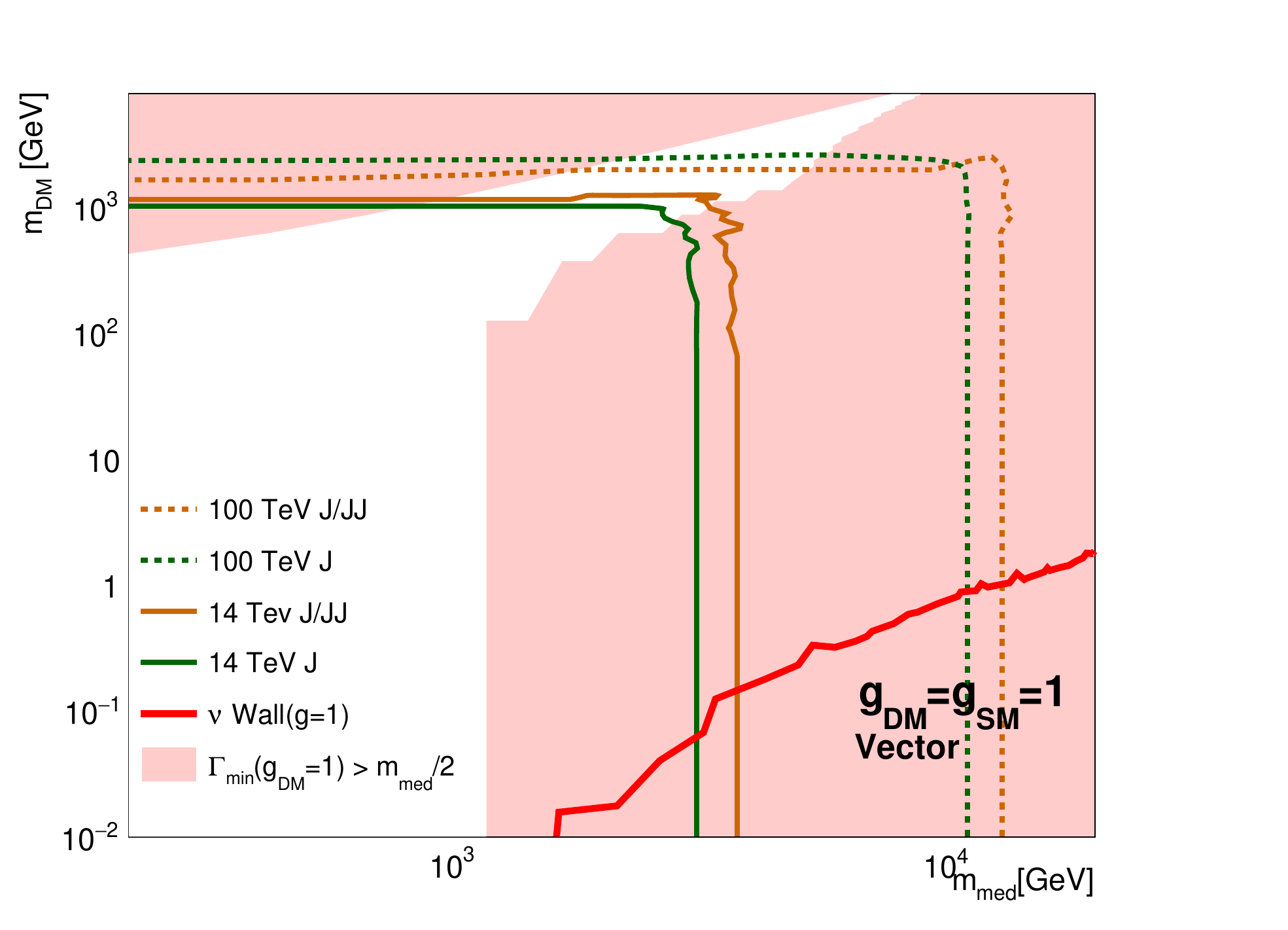} \hskip-0.6cm
\includegraphics[width=0.5\textwidth]{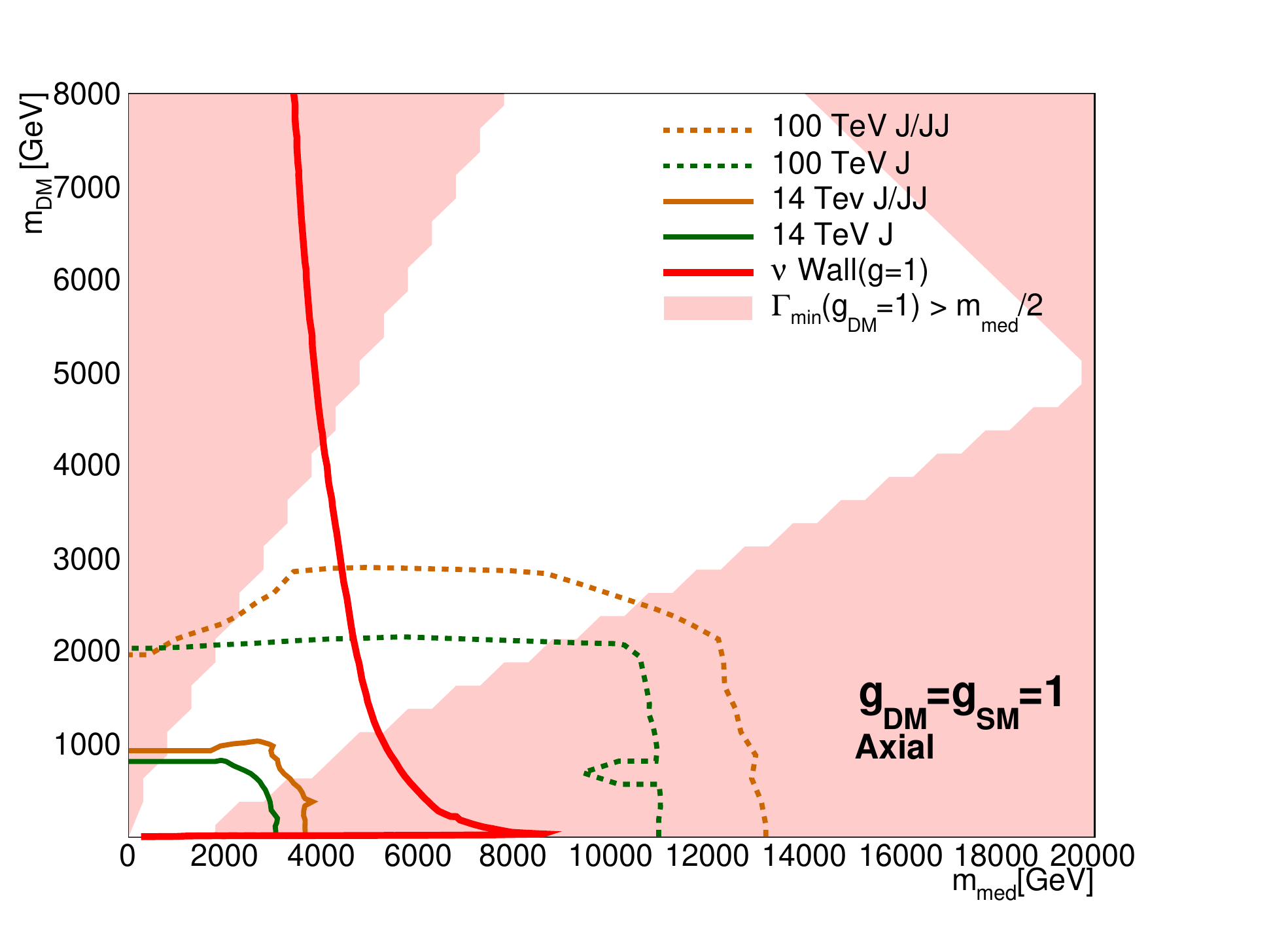}
\caption{Mass limits for vector mediator models (left panel) and axial-vector models (right pannel) at $14$ and $100$ TeV colliders 
using the multi-leg and a single-leg analysis. We also show the neutrino wall limit of the direct detection. }
\label{fig:exclMass:9}
\end{figure}
\end{center}

\begin{center}
\begin{figure}[h]
\includegraphics[width=0.5\textwidth]{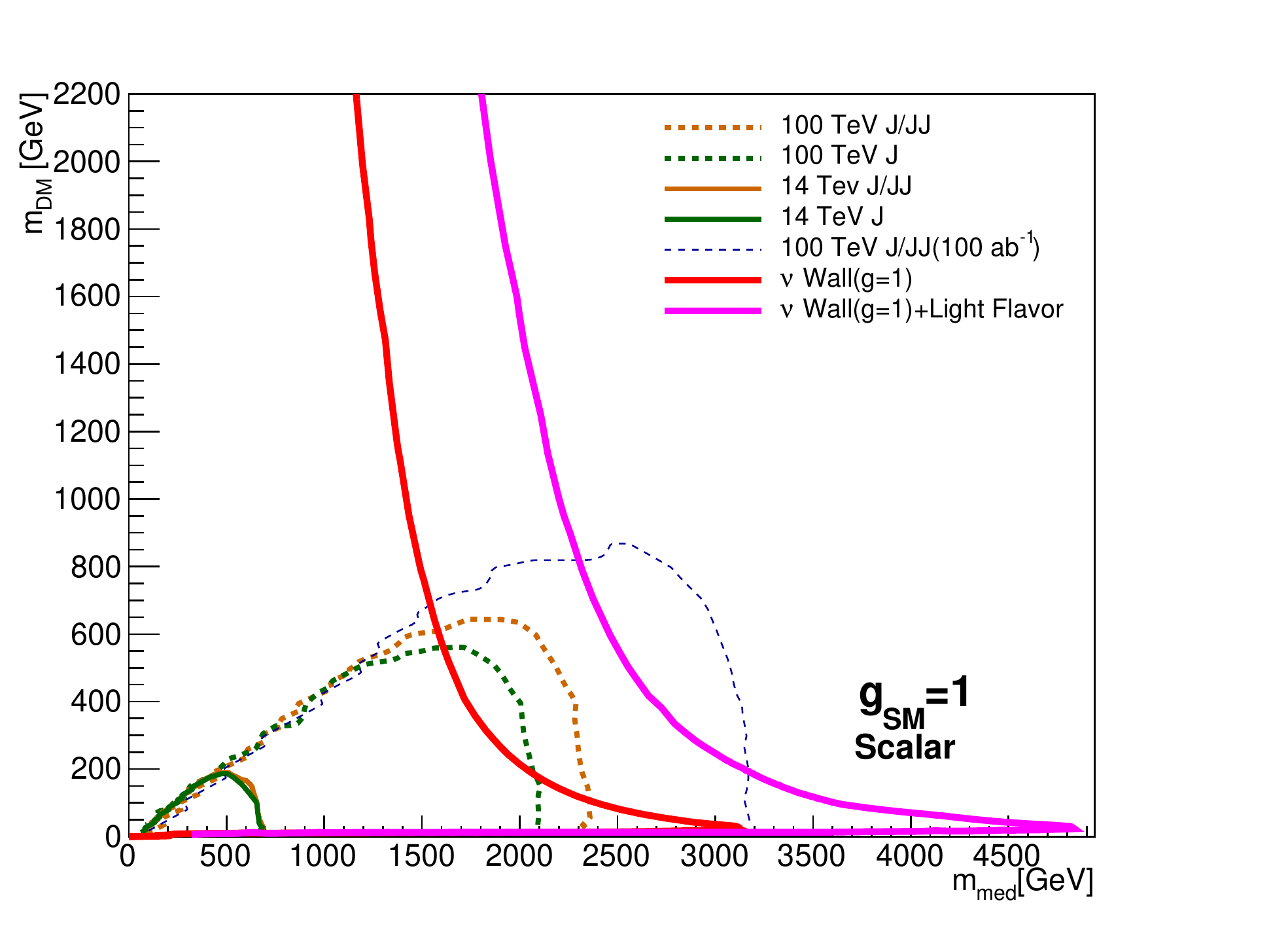} \hskip-0.6cm
\includegraphics[width=0.5\textwidth]{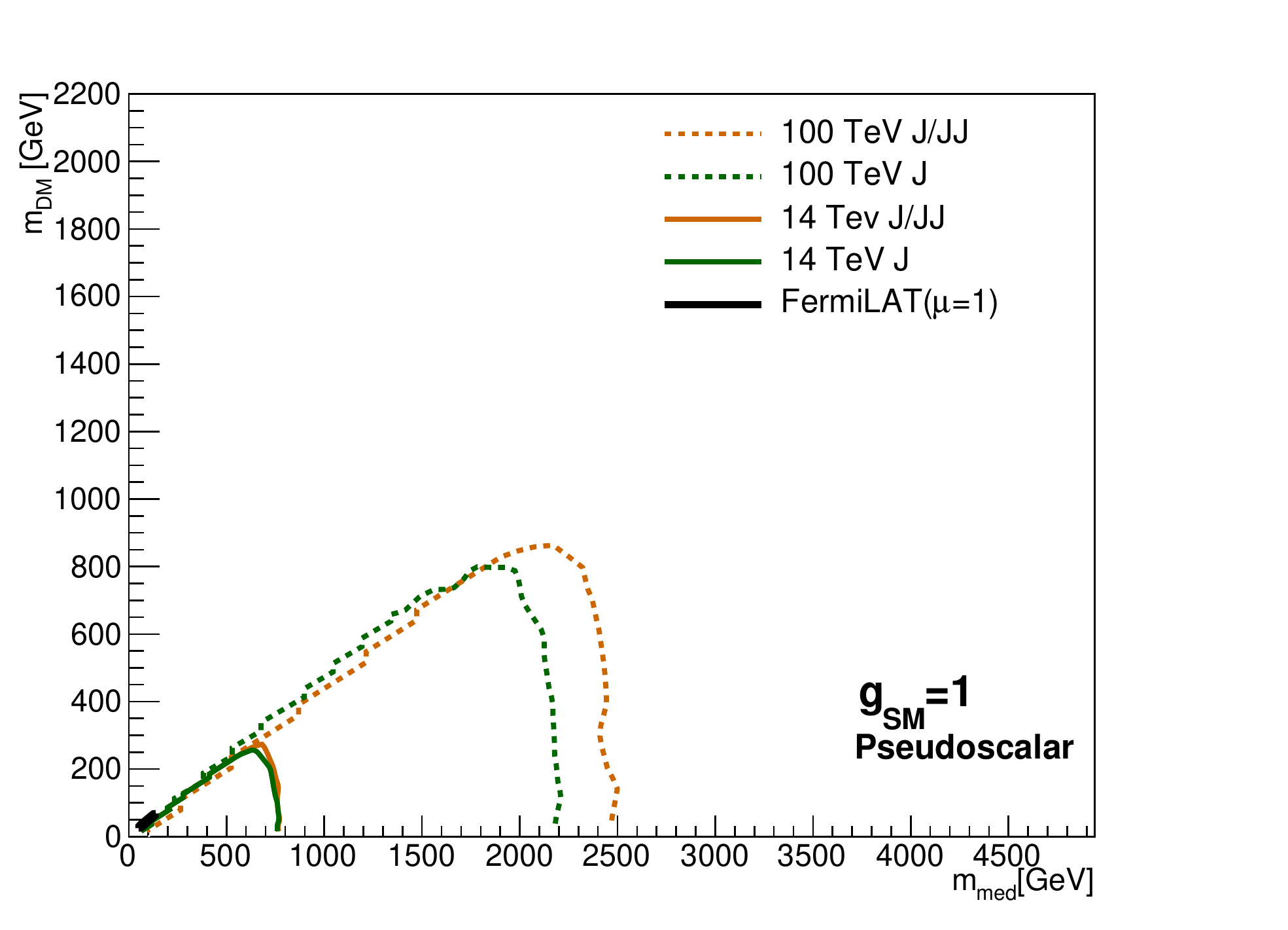}
\caption{Mass limits for scalar mediator models (left panel) and pseudo-scalar models (right pannel) at $14$ and $100$ TeV colliders 
using the multi-leg and a single-leg analysis. The neutrino wall affecting the direct detection experiments is shown in the left plot
and the indirect detection limit for pseudo-scalars using FERMI-LAT data~\cite{Ackermann:2011wa} is shown as a tiny speck in the lower left of the plot on the right.}
\label{fig:exclMass:10}
\end{figure}
\end{center}

\begin{center}
\begin{figure}[h]
\includegraphics[width=0.48\textwidth]{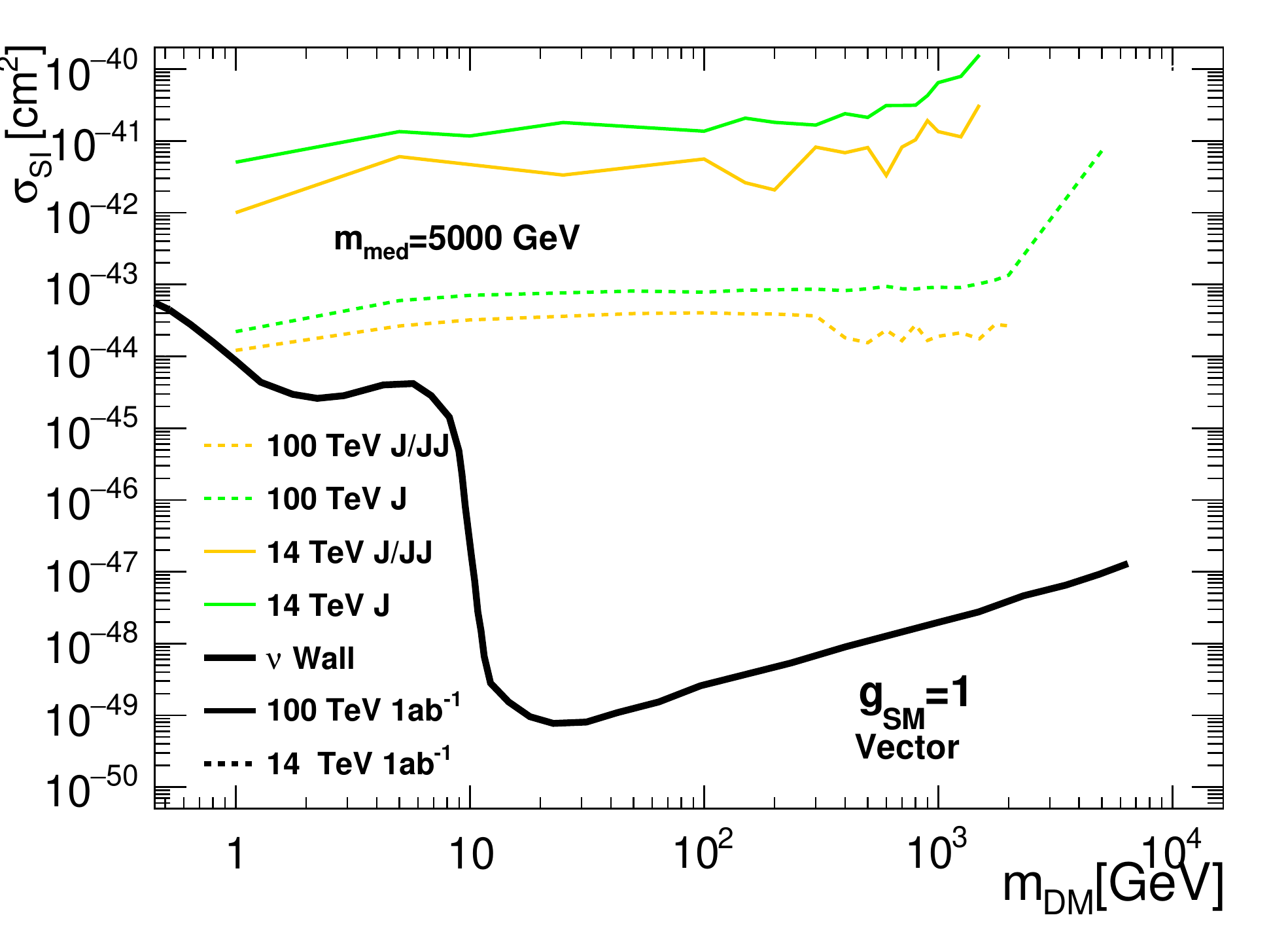}
\includegraphics[width=0.48\textwidth]{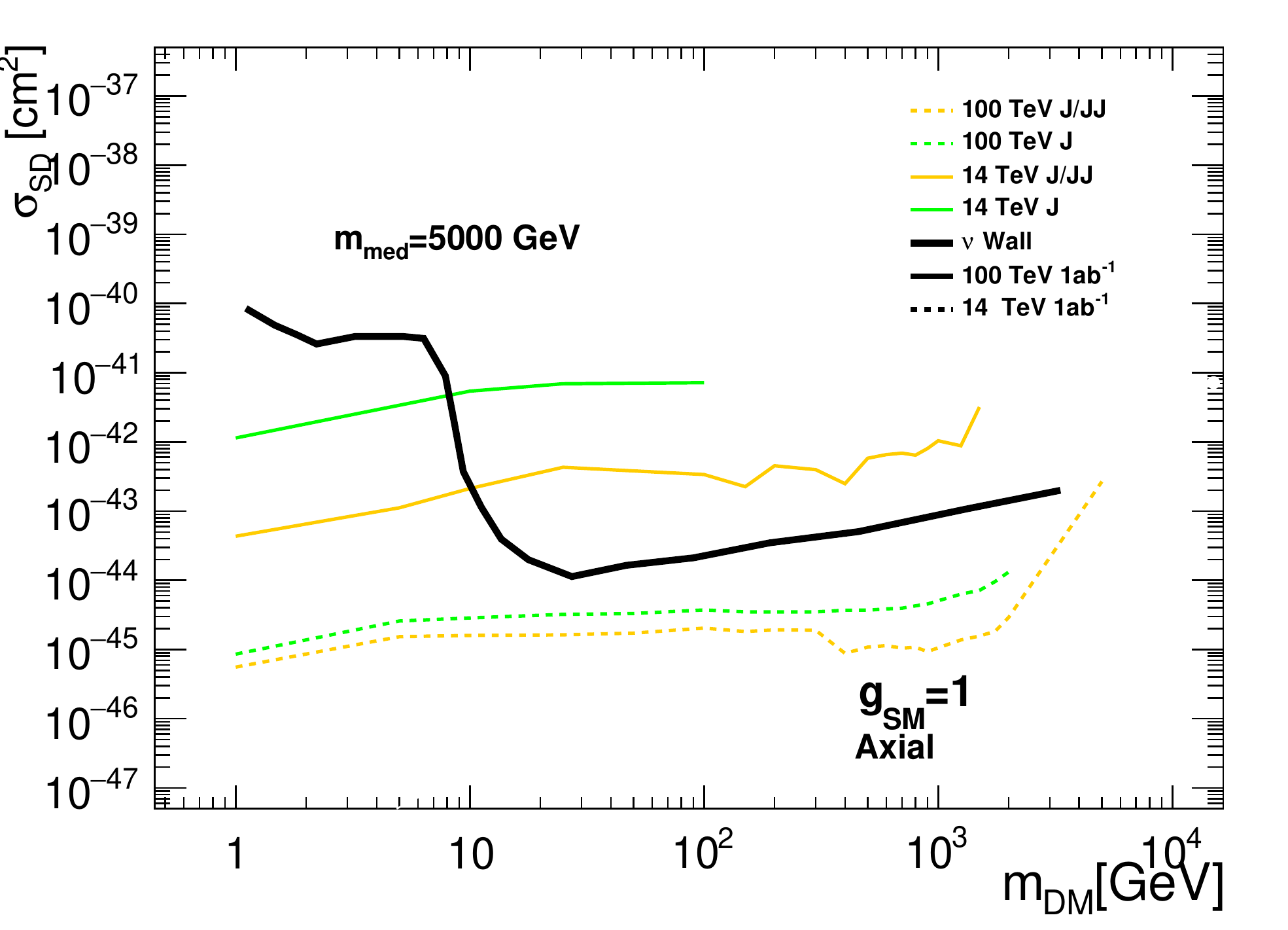}\\
\includegraphics[width=0.48\textwidth]{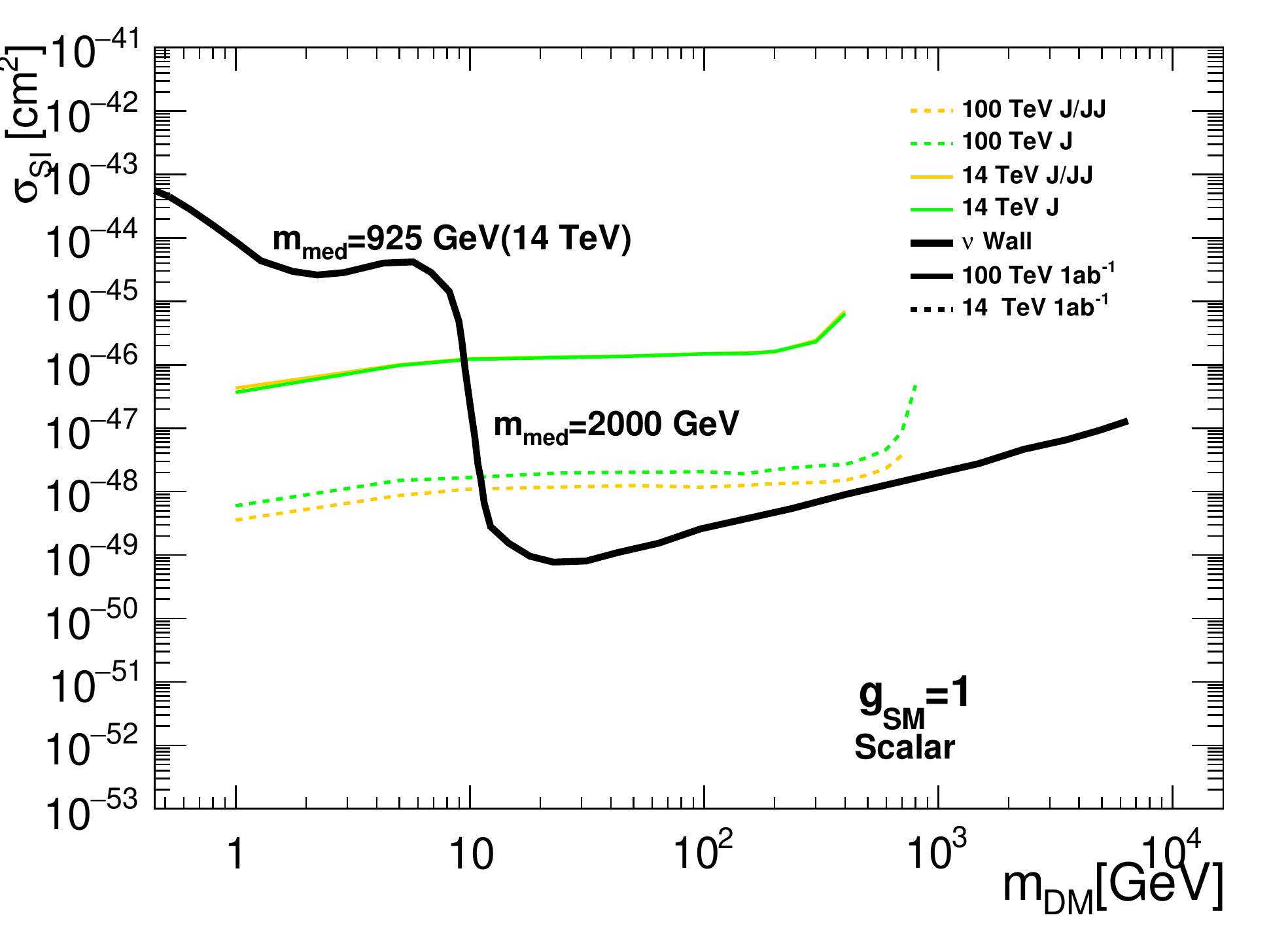}
\includegraphics[width=0.48\textwidth]{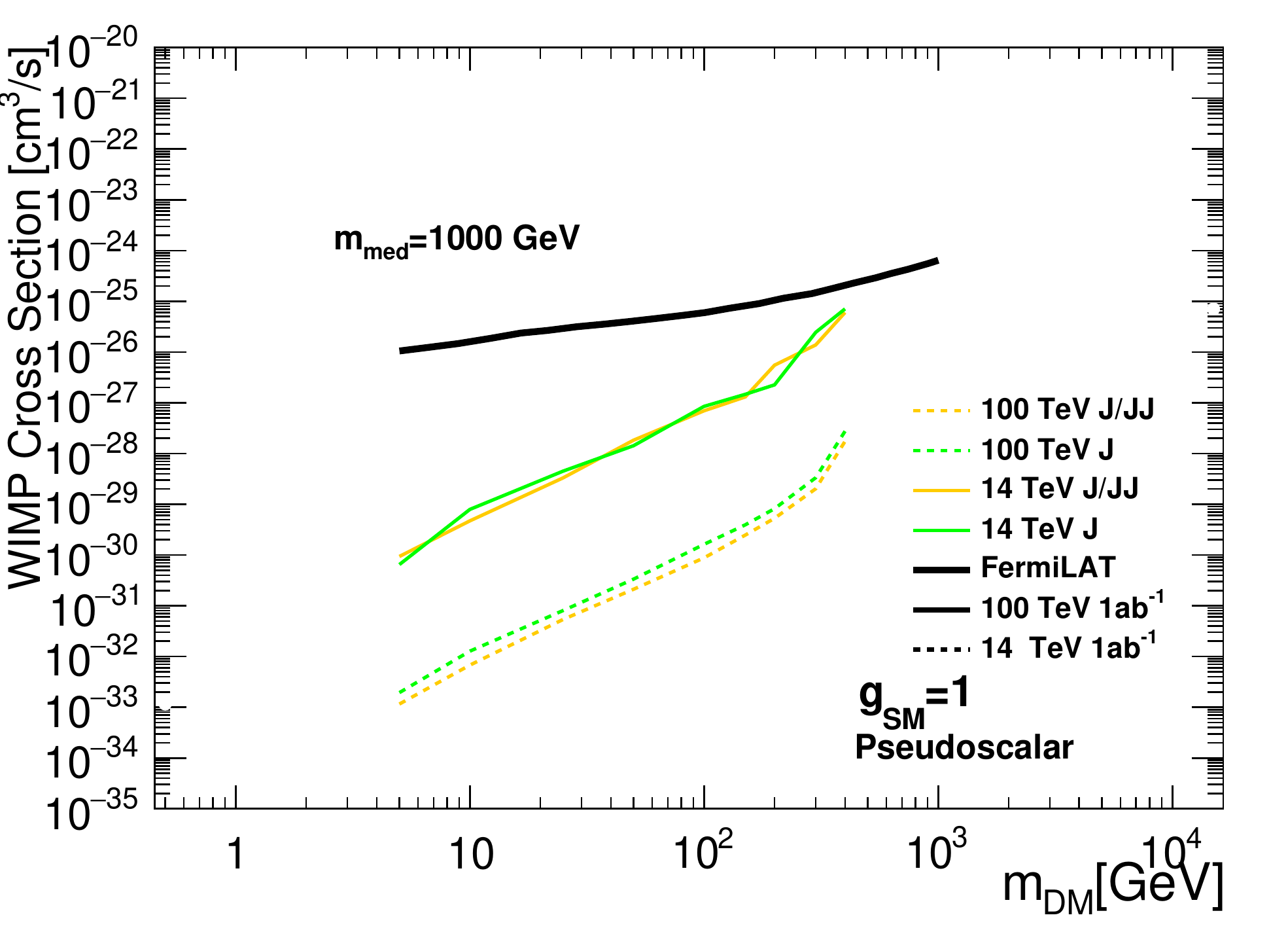}
\caption{Collider exclusion contours interpreted in terms of the spin-dependent and spin-independent cross sections 
plotted as functions of the dark matter mass. For the vector, axial-vector and scalar mediator models indicate the neutrino wall
limit of the direct detection experiments. For the pseudo-scalar mediator model we
show the indirect detection limits (using FERMI-LATdata~\cite{Ackermann:2011wa}). }
\label{fig:excl:11}
\end{figure}
\end{center}

 Figure~\ref{fig:excl14TeV_Mmed_mDM} shows the 14 TeV LHC exclusion limits for all four mediator models interpreted 
 on the ($m_{\rm MED}$, $m_{\rm DM}$) plane, and the color-coding indicates the value of the mediator coupling, assuming $g_{\rm DM}=g_{\rm SM}$.
 To provide a benchmark for the collider reach for searches of DM, we compare their projected exclusion contours with those
 relevant to the present and future direct and indirect detection experiments. Specifically, for the cases of scalar, vector and axial-vector
 mediators we show the neutrino wall which is expected to correspond to the ultimate reach of the direct detection (DD) experiments. 
 For the pseudo-scalar mediators one should compare with the indirect detection (ID) (to avoid the non-relativistic velocity suppression 
 affecting pseudo-scalar cross-sections in DD experiments). Hence for the pseudo-scalar mediators we
show the indirect detection limits using FERMI-LAT~data~\cite{Ackermann:2011wa}.
Figure~\ref{fig:excl100TeV_Mmed_mDM} generalizes the results to the FCC case with 100 TeV energy. We find large improvements in reach for the mediator mass $m_{\rm MED}$ as well as the dark matter mass $m_{\rm DM}$. Increasing the center-of-mass energy from 14 to 100 TeV allows us to probe 4-5 times larger mediator and dark matter masses, irrespective of the mediator's spin.
It is interesting to note that the limits obtained for the neutrino wall for vector, axial and scalar mediators included completely those 
of the LHC operating at 14 TeV. That is to say that, given enough time, data and understanding of astrophysical backgrounds, DD experiments 
can ultimately outperform the LHC for these mediator types (albeit with several model dependent caveats which render the results complementary to one anther in many regards). On the other hand, for the FCC large regions of parameter space can be accessed for the the axial and scalar mediator which cannot be accessed by DD experiments 
limited by the neutrino wall. This extension of the parameter space for these types of models is a significant benefit of the FCC since it is unlikely this parameter region will ever be reached by current generations of experiments. Finally we note that as the couplings grow larger the width of the mediator grows, eventually for large enough values of the coupling it can exceed $m_{\rm MED}/2$ and a particle interpretation of the mediator is no longer valid. We highlight these regions for the vector and axial vector case. For the scalar case, which corresponds to our multi-leg on-shell sample the limit is  trivially imposed by the on-shell condition (for the region where $m_{\rm DM} > m_{\rm MED}/2$). We therefore only present the contours for the cases where they are relevant.

In Figures~\ref{fig:exclMass:9} and \ref{fig:exclMass:10} we show these exclusion contours for the characteristic to a simplified model analysis
 fixed value of the mediator couplings, $g_{\rm DM}=g_{\rm SM}=1$ for all 4 mediator models of Eqs.~\eqref{eq:LS}-\eqref{eq:LA}. To enable the direct comparison between different experiments/techniques, these figures show all five exclusion contours -- 
 the 14 TeV and the 100 TeV limits, using both the one-jet and the multi-jet 
 analysis, together with the DD/ID non-collider limits/projections. 
 
 It is interesting to note the dependence of the DD limits in the scalar mediator case on the number of quark degrees of freedom it couples to.
 Unlike the production mechanism at collider searches which is sensitive only to the heavy top quark, the DD limits are sensitive also to
 light degrees of freedom thanks to the cancellation of the quark mass in the $y_q/m_q$ factor in Eq.~\eqref{eq:17}.
 Thus, the DD limits are quite sensitive to choice of flavors that mediator couples to in the simplified model.
 The magenta contour in in Figure~\ref{fig:exclMass:9}
 represents the inclusion of interactions with all quark flavors (as in the simplified model in Eq.~\eqref{eq:LS}).
 For a different choice of the simplified model, for example with only the top quark couplings to the mediator, the DD contour
 is shown in red. The difference between the red and magenta contours in the scalar mediator case in Fig.~\ref{fig:exclMass:9} 
 shows the sensitivity of the DD limits to a range of simplified models; at the same time the collider searches are are primarily sensitive to
 the scalar-to-top couplings\footnote{We note that in the previous figures the $\nu$-wall curve corresponds to the magenta curve.}. For this parameter choice we note that the collider constraints lie below the neutrino wall for 1 ab$^{-1}$, as the FCC collects more data the wall can be breached. As an example we plot the expected limit given 100 ab$^{-1}$ of FCC data for the scalar mediator.

Finally in Figure~\ref{fig:excl:11}
we show the plots in terms of the spin-dependent and the spin-independent DM--neutron cross sections 
for a more traditional comparison of collider limits in terms of our simplified models with the limits/projections  from 
the direct and indirect detection experiments. We compare the results in the 
$\sigma$, $m_{DM}$ plane. The general pattern of Figs.~\ref{fig:exclMass:9}-\ref{fig:exclMass:10} is reproduced, 
with the $\nu$-wall for the spin-independent cases providing the strongest projected bounds. For the axial-vector and
scalar mediators(with TeV-scale mediator masses as chosen in Fig.~\ref{fig:excl:11})
 our LHC contours cross the neutrino wall limits of direct detection experiments for $m_{\rm DM} \lesssim 10$ GeV.   At 100 TeV
 we find that collider bounds for the axial mediator are the strongest and universally below the $\nu$-wall limit of direct detection, 
 whilst for these parameter choices the scalar mediator and the $\nu$-wall are comparable. 
 In the pseudo-scalar case the last plot in Fig.~\ref{fig:excl:11} demonstrates that both 14 TeV and 100 TeV collider bounds 
 provide a multi-order of magnitude improvement over the current ID reach.

\section{Conclusions}
\label{sec:conc}

We have presented a comprehensive study of the forthcoming and future hadron collider limits and projections at 14 and 100 TeV
for searchers of new physics associated with the Dark Matter sector. The dark sectors are characterised in this work in terms
of four generic classes of simplified models 
Eqs.~\eqref{eq:LS}-\eqref{eq:LA}
where interactions between the Standard Model partons and the 
`invisible' dark matter sector particles are described by four basic types of mediators: scalar, pseudo-scalar, vector
or axial-vector particles. The dark matter particles we consider are produced via $s$-channel mediator exchange,
see Fig.~\ref{fig:feyn}. For collider searches of dark particles to be effective, two body decays of mediators produced on-shell 
should be kinematically possible, which requires that $m_{\rm MED} > 2 m_{\rm DM}$, as can be seen in e.g. 
Figs.~\ref{fig:excl14TeV_Mmed_mDM}.
It is then the ability to produce the mediator particle itself that underlines the efficiency of collider searches for dark matter,
and not so much the particular species of Dark Matter the mediator decays into, in so far as these decays are kinematically allowed.
Importantly, there is no requirement that once produced, the mediators should decay predominantly into the cosmologically stable Dark Matter,
instead (and arguably more plausibly) they can have significant branching ratios for decay into any dark sector particles which 
are long-lived on collider scales. Hence we do not impose the relic density constraints on the dark particle production in our simplified
model treatment.

Collider limits on the signal cross-sections for the cases of the 14 TeV LHC and the 100 TeV FCC are summarised 
in Fig.~\ref{fig:exclCrossS} for all four mediator types. From these we have determined the collider reach on the ($m_{\rm DM}$, $m_{\rm MED}$) mass plane at 14 TeV in Fig.~\ref{fig:excl14TeV_Mmed_mDM} and 100 TeV in Fig~\ref{fig:excl100TeV_Mmed_mDM}. 
Both these sets of figures compare the collider reach with the 
neutrino wall limit of direct detection experiments and the current data from the indirect detection (in the latter case for the pseudo-scalar
mediator models). We conclude that for scalar and axial-vector mediators collider searches are highly competitive and also complimentary to 
direct detection experiments, while in the pseudo-scalar case the collider limits are unchallenged. These figures also show the required strength 
of the DM coupling of the mediators, which in all cases remains largely in the perturbative regime. 

Figures \ref{fig:exclMass:9} and \ref{fig:exclMass:10} combine the 14 TeV and 100 TeV projections 
for collider searches and compare them with DD and ID limits
at fixed values of mediator couplings $g_{\rm DM}=g_{\rm SM}=1$. For vector and axial-vector mediators mediator masses of up to 
$\sim 3$ TeV can be probed at the LHC and extended to nearly 13 TeV at the 100 TeV FCC. For scalar and pseudo-scalar mediators,
the 14 TeV LHC reach is of $m_{\rm MED} \sim 700$ GeV and up to $\sim$ 4-5 TeV for the FCC. 
Thus the Future hadron Circular Collider would be able to truly probe the few-TeV scales of the Dark Matter sectors, and in this
regime it crosses over and goes beyond the projected limits of DD and ID experiments as can be seen from 
Figs.~\ref{fig:exclMass:9}-\ref{fig:exclMass:10}
on the $m_{\rm DM}$, $M_{\rm MED}$ mass plane
and the plots in Fig.~\ref{fig:excl:11} which interpret collider searches in the language of DD experiments in terms of the spin-independent and spin-dependent cross-sections.
Our FCC limits were presented assuming a rather modest data set of 1 ab$^{-1}$, over the lifetime of the machine a much larger data set should be collected, and the 
limits/discovery sensitivity will be significantly enhanced. 

In summary, although our results for the FCC are somewhat speculative, it is clear that there is a huge potential for such a machine to probe a large parameter space 
for a variety of dark sector scenarios. Indeed an FCC style machine can provide access to regions of parameter space for axial and scalar mediators which lie beyond the potential limits of the cosmic neutrino wall. Such a possibility is an exciting aspect of the emerging physics program of the FCC, and in our opinion represents a strong motivation to press ahead with its ultimate construction. 

\section*{Acknowledgments}
\noindent We would like to thank Oliver Buchmueller, Krisitan Hahn,
Michelangelo Mangano, Christopher McCabe, Filip Moortgaart, Nick
Wardle, and Nhan Tran, for valuable discussions.
The research of VVK and MS is supported by STFC through the IPPP grant and for VVK by the Wolfson Foundation and Royal Society.
VVK acknowledges the Aspen Center for Physics supported by National Science Foundation grant PHY-1066293, where this work was finalized.

\bibliography{ref}
\bibliographystyle{ArXiv}

\end{document}